\documentclass[sigconf]{acmart}
\renewcommand\footnotetextcopyrightpermission[1]{} 
\usepackage{tugraz_defaults}
\usepackage{balance} 
\graphicspath{{figures/}}
\DeclareGraphicsExtensions{.pdf,.jpeg,.png}

\begin{document}
\fancyhead{}
\newcommand{\riscv}{\mbox{RISC-V}\xspace}
\newcommand{\sanc}{\texttt{HECTOR-V}\xspace}
\newcommand{\cpu}{\texttt{RVSCP}\xspace}

\def\ci#1{\textcircled{\resizebox{.5em}{!}{#1}}}
\makeatletter
\renewcommand\paragraph{\@startsection{paragraph}{4}{\z@}%
                                      {\parskip}
                                      {-0.5em}%
                                      {\normalfont\normalsize\bfseries}}

\title{\sanc: A Heterogeneous CPU Architecture for a Secure RISC-V Execution Environment}

\author{Pascal Nasahl}
\email{pascal.nasahl@iaik.tugraz.at}
\affiliation{%
  \institution{Graz University of Technology}}

\author{Robert Schilling}
\email{robert.schilling@iaik.tugraz.at}
\affiliation{%
  \institution{Graz University of Technology}}

\author{Mario Werner}
\email{mario.werner@iaik.tugraz.at}
\affiliation{%
  \institution{Graz University of Technology}}

\author{Stefan Mangard}
\email{stefan.mangard@iaik.tugraz.at}
\affiliation{%
  \institution{Graz University of Technology}
  \institution{Lamarr Security Research}}

\begin{abstract}
To ensure secure and trustworthy execution of applications in potentially insecure environments, vendors frequently embed trusted execution environments~(TEE) into their systems.
Applications executed in this safe, isolated space are protected from adversaries, including a malicious operating system.
TEEs are usually build by integrating protection mechanisms directly into the processor or by using dedicated external secure elements.
However, both of these approaches only cover a narrow threat model resulting in limited security guarantees.
Enclaves nested into the application processor typically provide weak isolation between the secure and non-secure domain, especially when considering side-channel attacks.
Although external secure elements do provide strong isolation, the slow communication interface to the application processor is exposed to adversaries and restricts the use cases.
Independently of the used approach, TEEs often lack the possibility to establish secure communication to peripherals, and most operating systems executed inside TEEs do not provide state-of-the-art defense strategies, making them vulnerable to various attacks.

We argue that TEEs, such as Intel SGX or ARM TrustZone, implemented on the main application processor, are insecure, especially when considering side-channel attacks.
In this paper, we demonstrate how a heterogeneous multicore architecture can be utilized to realize a secure TEE design.
We directly embed a secure processor into our \sanc architecture to provide strong isolation between the secure and non-secure domain.
The tight coupling of the TEE and the application processor enables \sanc to provide mechanisms for establishing secure communication channels between different devices.
We further introduce \riscv Secure Co-Processor~(\cpu), a security-hardened processor tailored for TEEs.
To secure applications executed inside the TEE, \cpu provides hardware enforced control-flow integrity and rigorously restricts I/O accesses to certain execution states.
\cpu reduces the trusted computing base to a minimum by providing operating system services directly in hardware.
\end{abstract}


\maketitle

\section{Introduction}
\label{sec:intro}
With the growing demand for complex IT applications, such as autonomous driving or smart city infrastructures, software complexity increases steadily.
The codebase of Linux, for example, increases by 250k lines of code each year, reaching 27.8M LOC in 2020~\cite{LinuxLOC}.
This is challenging because one can expect roughly 1 to 25 bugs in 1,000 lines of code~\cite{mcconnell2004code}.
While not all of these bugs might be exploitable by an attacker, a growing codebase complexity clearly leads to a larger attack surface.
One strategy to deal with the growing complexity and meet security goals is to isolate all security critical applications using a trusted execution environment~(TEE)~\cite{DBLP:journals/ieeesp/EkbergKA14}.

A TEE establishes a secure execution environment by creating a safe, isolated space using hardware and software primitives.
Most TEE threat models consider a powerful adversary controlling user space applications, the operating system, or even the hypervisor, trying to influence the execution of applications in the trusted environment.
To achieve the protection needed for this threat model, TEEs require strong isolation between the TEE and the rich execution environment~(REE).

The powerful concept of shifting security critical applications into a trusted execution environment is already adapted by major vendors like Intel, ARM, and Apple.
One of the most common TEE design approaches is to create a virtual secure processor in the main application processor by using hardware extensions.
Intel SGX~\cite{SGX2019} and ARM TrustZone~\cite{arm2009security} take this approach.
Contradictory to embedding the TEE tightly into the CPU, Google's Titan~\cite{johnson2018titan} and Apple's T2~\cite{AppleT2} implements a secure element by externally mounting a dedicated security processor next to the main CPU.
However, both TEE design approaches yield different weaknesses.
Several recent attacks~\cite{DBLP:conf/eurosp/ZhangSSLH16, DBLP:conf/sp/GuancialeNBD16, DBLP:conf/uss/LippGSMM16,DBLP:conf/woot/BrasserMDKCS17,DBLP:conf/dimva/SchwarzWGMM17,DBLP:conf/eurosec/GotzfriedESM17,DBLP:conf/uss/BulckMWGKPSWYS18,van2020lvi, DBLP:journals/ieeesp/ChenCXZLL20} showed that the isolation of TrustZone and SGX can be bypassed by mounting cache or transient-based side-channel attacks.
While dedicated secure elements provide strong isolation between REE and TEE, Google's Titan, \eg uses a slow SPI connection for communication between the two domains limiting potential use cases.
Furthermore, this off-chip communication fabric between REE and TEE is physically exposed to an attacker, making it vulnerable against probing attacks.

Independently of the used TEE design approach, typical TEE implementations offer several weaknesses.
First, although a security breach within the TEE is fatal, operating systems deployed in the secure environment surprisingly do not offer state-of-the-art protection mechanisms like ASLR, guard pages, or stack cookies~\cite{cerdeira2020sok}.
Second, most TEEs do not provide architectural features to establish secure communication with I/O devices.
The lack of trusted I/O paths in TEE systems is critical because secrets shared between user and TEE are unprotected.

With the Intel Lakefield architecture~\cite{DBLP:conf/hotchips/KhushuG19} and an AMD patent~\cite{AmdPatent}, major vendors recently announced to introduce heterogeneous multi-core architectures in upcoming processor designs.
While the approach of tightly coupling different processor cores on one chip to balance power and energy efficiency is already used by ARM's big.LITTLE technology~\cite{ARMbigLittle} in mobile platforms, Intel and AMD are planning to introduce this concept in forthcoming computer architectures.
This design strategy raises the following research question: 
\textit{RQ1: "Can the tight coupling of distinct processor cores on a SoC be used to increase the security of trusted execution environments?"}
\subsection*{Contribution}

In this paper, we introduce \sanc, a design methodology utilizing a heterogeneous multi-core architecture to develop secure trusted execution environments.
\sanc achieves strong isolation between the secure and non-secure domain and provides protection for various side-channel attacks, such as cache and transient-based attacks, by using a distinct processor of the heterogeneous core cluster for the secure environment.
The tight coupling of TEE and REE in the shared SoC infrastructure yields several advantages.
In the \sanc architecture, the application processor and TEE are connected through an interconnect, enabling a high-speed link between the two processors.
Additionally, since the TEE is embedded into the SoC, all peripherals integrated into the system are also available for the secure environment.
To manage peripheral sharing and protect peripherals from unauthorized accesses, \sanc further introduces a secure I/O path concept.
The identifier-based strategy deeply nested into the communication fabric and the peripherals allows a fine granular protection of the attached peripherals.
In \sanc, the access permission management is implemented using a hardware-based security monitor.
While the owner of the security monitor can configure a set of access permissions, the other parties in the system can request access to certain peripherals by consulting the security monitor.
We further extend this mechanism to dynamically grant and deny access to certain peripherals by introducing the concept of security monitor ownership transfer.
To enable various use cases, the owner of the security monitor dynamically can transfer the ownership to other parties.
Furthermore, we introduce \cpu, a security-hardened \riscv processor tailored for being used as a TEE. 
Although the \sanc architecture is independent of the used TEE processor, we show that \cpu further increases the TEE security by utilizing features of \sanc.
\cpu isolates applications within the trusted execution environment by enforcing the integrity of the control-flow.
We combine the control-flow information with the secure I/O mechanism to only grant access to certain peripherals when reaching a predefined control-flow state.
To reduce the trusted computing base and therefore the attack surface to a minimum, \cpu implements operating system features, such as multitasking, directly in hardware.
We demonstrate the \sanc concept by introducing a heterogeneous multi-core architecture for \riscv.
We embedded a \riscv processor with a control-flow integrity unit into the \sanc architecture.
To verify the functionality of \sanc and \cpu design approaches, we use secure boot as a prototype application on our FPGA implementation.
In summary, our contributions are:
\paragraph{\sanc,} a design methodology for developing secure TEEs.\\
\sanc utilizes a heterogeneous multi-core architecture to realize a secure trusted execution environment.
The architecture includes a configurable security monitor managing access permissions to specific peripherals.
This mechanism allows the system to establish secure communication channels between peripherals, users, and the processor cores.
The dynamic transfer of security monitor ownership enables the \sanc architecture to realize several use cases, such as secure boot or executing trusted applications.
\paragraph{\cpu,} a security-hardened processor integrated into the \sanc architecture.
\cpu protects applications within the TEE by combining a previously introduced control-flow integrity unit with the peripheral access protection offered by \sanc.
The secure processor provides operating system features in hardware to minimize the trusted computing base.
\section{Background}
\label{sec:background}

To protect the processing of sensitive data, such as biometric data or cryptographic keys in an untrusted environment, trusted execution environments (TEE) are used.
These secure environments comprise a combination of hardware and software features, the trusted computing base (TCB)~\cite{DBLP:journals/ieeesp/EkbergKA14}.
A TEE guarantees the secure execution of trusted applications, even when considering a malicious operating system running in the rich execution environment~(REE).
This section summarizes popular TEE design strategies.

\paragraph{Virtual Processor based TEEs.}
Virtual processor based TEEs extend the main processor with additional hardware and software features to establish a secure, isolated space within this processor.
As this approach utilizes the main processor for the TEE, such TEEs achieve high performance and require no additional hardware~\cite{DBLP:journals/ieeesp/KostiainenDC20}.
Hence, major vendors, like Intel with SGX~\cite{SGX2019} and ARM with TrustZone~\cite{arm2009security}, utilize this approach.
However, sharing resources, such as caches or peripherals, also is the main disadvantage of such systems.
SGX and TrustZone fail to provide strong isolation guarantees by being vulnerable to side-channel attacks, such as cache attacks~\cite{DBLP:conf/eurosp/ZhangSSLH16, DBLP:conf/sp/GuancialeNBD16, DBLP:conf/uss/LippGSMM16,DBLP:conf/woot/BrasserMDKCS17,DBLP:conf/dimva/SchwarzWGMM17,DBLP:conf/eurosec/GotzfriedESM17} or transient attacks~\cite{DBLP:conf/uss/BulckMWGKPSWYS18,van2020lvi, DBLP:journals/ieeesp/ChenCXZLL20}.
Additionally, most TEEs, \eg SGX, based on this design approach, do not protect peripherals by providing secure I/O.

\paragraph{External Co-Processor based TEEs.}
These TEEs utilize a dedicated, external core to securely execute trustlets.
Examples of such secure elements are Google's Titan~\cite{johnson2018titan} and Samsung's eSE~\cite{ese}, which are mainly used in mobile platforms or servers.
Since this technique separates the execution pipelines and the caches of REE and TEE, cache- and transient-based side-channel attacks are mitigated efficiently.
Although these TEEs provide strong security guarantees, they also have major drawbacks.
External secure elements typically establish a communication channel between REE and TEE using a slow communication interface, like SPI.
This limits potential use cases of the TEE and the exposed interface also is vulnerable against physical probing and sniffing attacks~\cite{bussniffing}.
Furthermore, dedicated secure co-processors require additional hardware and do not support shared peripherals.

\paragraph{On-SoC Processor based TEEs.}
This approach combines advantages from virtual processor and external co-processor based TEE designs.
Here, a dedicated secure element is directly embedded into the main SoC infrastructure of the system.
Although placing such an element into the SoC increases the die size of the chip, this approach enables fast communication channels between REE and TEE and aggravates probing and sniffing attacks on this interface.
Therefore, and due to the possibility to share peripherals on the SoC, this approach increases the flexibility and possible use cases of the design. 
Additionally, by placing two separate cores into the system, cache- and transient-based side-channel attacks can be mitigated.
This approach is taken by major vendors, like Apple with their Secure Enclave Processor~(SEP)~\cite{seppatent} and Microsoft with their recently announced Pluton processor~\cite{pluton}.
\section{Threat Model}
\label{sec:adversary_model_design_goals}

Our threat model considers the common attack scenarios on TEEs defined by~\citeauthor{cerdeira2020sok}~\cite{cerdeira2020sok}.
We are considering an attacker directly exploiting architectural weaknesses of the TEE and the TEE kernel.
Here an adversary uses a combination of bugs in the kernel, flaws in the hardware protection mechanism, and missing state-of-the-art defense strategies, such as ASLR or guard pages, to compromise the system.
Furthermore, we expect bugs in trustlets, which can be exploited by an attacker over the communication interface between REE and TEE.
The security of applications in REE or TEE also can be threatened by using cache-based or transient-based side-channel attacks.
Additionally, we are considering a malicious trustlets explicitly trying to attack the SoC.
We are extending the threat model of common TEEs also to cover attacks on peripherals, \ie illegitimate accesses to secure storage elements, sensitive peripherals such as a fingerprint reader or SPI~\cite{CVESpi}, or protected memory regions.
In summary, we expect a powerful attacker having full control over user applications, the operating system, trustlets, or even the hypervisor executed on the application processor.

\section{Design}
\label{sec:design}

This section presents our secure TEE design approach consisting of the \sanc architecture and the secure processor \cpu.
We first introduce \sanc consisting of several architectural features, such as trusted I/O paths, a security monitor, and a secure TEE integration.
Then, we introduce \cpu, a concrete proposal for a secure processor used as a TEE and demonstrate, how \cpu and \sanc combined, form a secure TEE system.
In Section~\ref{sec:implementation}, we then demonstrate a prototype of \sanc and \cpu integrated into our \riscv base platform, the lowRISC chip.
\subsection{\sanc Design}

\begin{figure}[t]
  \begin{center}	
    \includegraphics[width=0.9\linewidth]{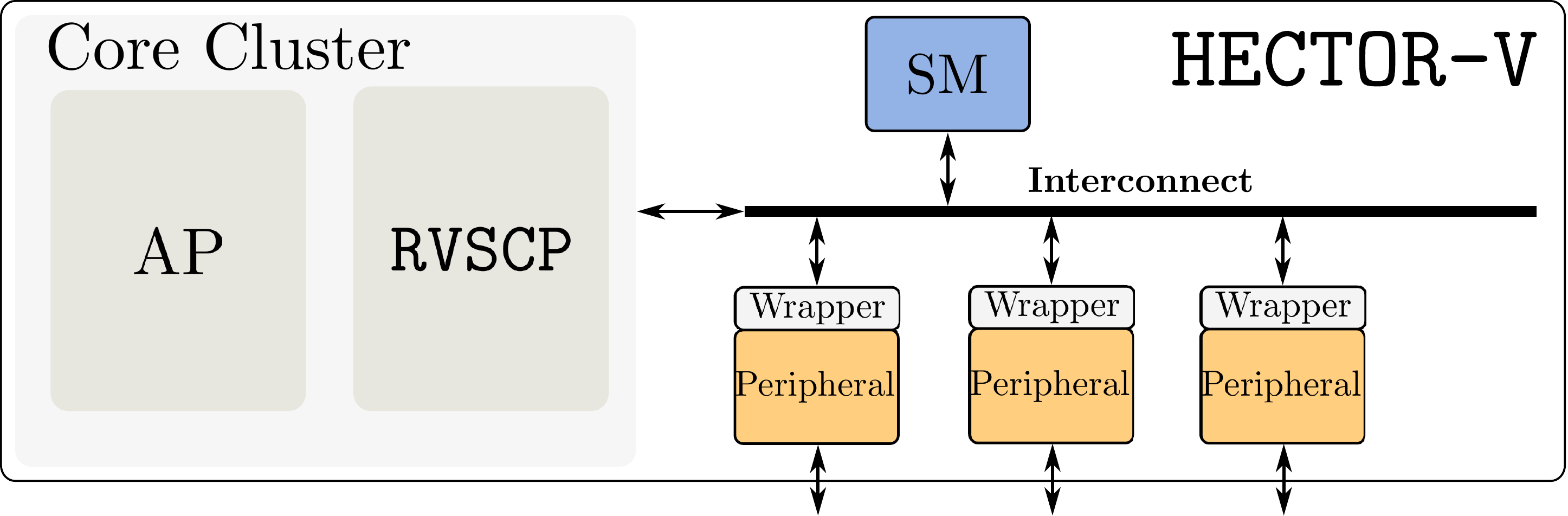}
    \caption{Overall \sanc design.}
    \label{fig:design}
  \end{center}
\end{figure}

The \sanc design proposes architectural features for creating secure TEEs.
As seen in Figure~\ref{fig:design}, the TEE in \sanc is realized by mounting a secure processor directly into the main SoC.
This approach is similar to ARM's big.LITTLE technology~\cite{ARMbigLittle}, where independent cores are embedded into the chip.
However, instead of using the additional cores for power efficiency, \sanc uses the dedicated processor for security purposes.
Although placing a secure processor directly into the SoC increases the area overhead, we argue that this approach is feasible and yields several advantages.
First, we contend that offering strong isolation between secure and non-secure domain only is possible with a dedicated secure processor for the TEE.
TEE architectures introduced before, such as Arm TrustZone and Intel SGX, are vulnerable to side-channel and microarchitectural attacks~\cite{DBLP:conf/uss/LippGSMM16, DBLP:journals/corr/abs-1802-09085}.
As these attacks are the result of the growing complexity of modern processors driven by the demand for high-speed systems~\cite{DBLP:journals/ieeesp/KostiainenDC20}, fixing these security issues in the main processor is challenging~\cite{miscflaw}.
However, the design choice of using a dedicated secure processor for the TEE completely separates the instruction pipelines and caches of the cores, resulting in an independent execution flow of REE and TEE applications.
Second, services deployed on TEE processors, such as applications handling banking information or processing passwords, are typically rather small.
This constrained processing requirement allow us to deploy a tiny secure processor.
Compared to a state-of-art multicore processor, the area overhead introduced by this tiny processor is negligible. 
While integrating a dedicated secure processor into the SoC infrastructure is straightforward, a secure and viable TEE system requires to establish communication channels between REE and TEE, as well as for peripherals enabling user I/O.
In the remaining part of this section, we introduce our trusted I/O path design utilizing a hardware-based security monitor and elaborate how these concepts allow REE and TEE to share devices on the SoC and realize features, such as secure storage elements.
\subsubsection{Trusted I/O Paths}
The trusted I/O path mechanism is the central element of the \sanc architecture.
This concept allows \sanc to securely share devices on the SoC, establish secure communication channels between external users and the TEE or REE, and implement concepts, like secure storage elements.
\sanc establishes trusted I/O paths by assigning an unique identifier to each party of the system, \ie the processor cores.
When accessing a device, like the SD-card or the block RAM~(BRAM), the peripheral uses an internal mechanism to verify the identifier.
This strategy allows the architecture to enforce access permissions, such as the fingerprint reader only can be accessed by the TEE.
\paragraph{Identifier.}
In \sanc, an identifier is used to distinguish between legitimate and illegitimate accesses to a peripheral.
While this concept is similar to ARM TrustZone~\cite{arm2009security}, TZ only uses a 1-bit identifier to distinguish between accesses from the secure or non-secure world.
The identifier used by our system consists of a core ID, a process ID, and a peripheral ID.
While the core ID permanently and unchangeable is assigned to each processing core, \eg an ID for the application processor and one ID for the secure processor, at design time, the process and peripheral ID can be assigned by each entity itself.

\paragraph{Interconnect.}
To efficiently transport the ID from the participants to the peripherals, we integrate the ID directly into the system interconnect.
The AMBA AXI4~\cite{axi} protocol, which is used as interconnect by our lowRISC base platform and many other SoC designs, allows to embed up to 1024-bits wide user-defined signals into the protocol.
By embedding the ID into the user-defined signals, which are not used by most AXI devices, the ID is transported without any protocol overhead.
Additionally, this approach assures that the identifier is sent along with the address and data on each AXI request.
Since the core ID is hardcoded directly into the the interconnect interface of the participants, an attacker cannot change this ID from software.

\paragraph{Peripherals.}

\sanc creates a trusted channel between an entity, \eg the application or the secure processor, and a peripheral by enforcing the identifier-based access check directly in hardware at the peripheral driver.
The peripherals, \eg the hardware implementation of the SD-card controller or the BRAM, use the ID to filter illegitimate accesses.
This scheme, which requires the peripherals to be identifier-aware, is implemented by introducing a lightweight wrapper module for each peripheral instance.
In our architecture, the peripheral wrapper module consists of two communication channels: the data channel and the configuration channel.
The data channel interface of the peripheral wrapper is directly connected to the AXI4 communication fabric, allowing other parties, \eg the secure processor, to access the peripheral.
By using the configuration channel, the configuration party can set or unset the identifier in the ID field.
In \sanc, the data and configuration channels are physically separated by introducing a dedicated, lightweight AXI4-lite~\cite{axi} communication fabric.

When the identifier of the AXI4 communication request transported over the communication fabric matches the identifier stored in the peripheral wrapper ID field, the access is permitted and the request is directly forwarded to the actual peripheral driver.
On an ID mismatch, the firewall mechanism blocks the request and returns an exception over the AXI4 channel, which can be processed by the issuer.
If the process or peripheral identifier is set to zero, the peripheral wrapper ignores these ID fields.

The peripheral wrapper resides in two states: claimed or unclaimed.
While in the unclaimed state all data channel requests are blocked, in the claimed state only the requests with the matching identifier are permitted.
Furthermore, we differentiate between configurable and non-configurable peripheral wrappers.
To realize functionalities, like secure storage elements, which must not be accessible by any party except one, non-configurable peripheral wrappers store a hardcoded identifier, consisting of the core and process ID, in the identifier field.
\subsubsection{Security Monitor}
The architectural features of identifier, identifier transportation, and peripheral firewalls are the foundation for establishing trusted I/O paths on the SoC.
To enforce security policies utilizing the trusted paths, we embed a tiny security monitor into the SoC infrastructure.
This hardware-based security monitor~(SM) module acts as trusted computing base~(TCB) and is responsible for managing access permissions to peripherals.
Internally, the SM consists of a table tracking these access permissions.
For each peripheral included in the system, the security monitor maintains a table entry tracking the state (claimed or unclaimed) and a list of identifiers allowed to claim the device.

\paragraph{Security Monitor Owner.}
The \sanc architecture introduces the security monitor owner privilege, which is assigned to a certain party at design time.
This party solely is responsible for configuring the TCB, \ie the security monitor, over an interface.
More concretely, only the SM owner is allowed to define, which peripheral can be claimed by which participant.
The security monitor stores the identifier of the current owner and accepts configuration commands only from this party.

\paragraph{Peripheral Claiming and Releasing.}
To access a certain peripheral, the requester first needs to claim the peripheral by sending a claim request to the security monitor.
Then, the SM checks, if the peripheral currently is unclaimed and verifies that the ID of the issuer is in the list of identifiers permitted to access the peripheral.
Finally, the security monitor sends the ID of the issuer to the peripheral wrapper over the AXI4-lite bus and this wrapper sets this identifier in its ID field.
Now, the identifier transported in the user-defined signals of each AXI4 request issued by the requester matches the identifier in the ID field and the requester exclusively can access the peripheral.
However, if the requester is not allowed to claim the peripheral, \ie the identifier of the requester is not in the list of privileged entities, the ID verification fails and the security monitor sends back an access denied message.  
If the requester ID is in the table entry of the peripheral, but the peripheral currently is claimed, the security monitor notifies the issuer.
\sanc is a cooperative scheme, meaning, the entity currently claiming a peripheral needs to release it after using it.
To release a peripheral, the entity sends a release command to the SM.
The security monitor processes this request by clearing the ID field of the peripheral wrapper.

\paragraph{Peripheral Access Withdraw.}
Claiming a peripheral and binding access exclusively to an entity is a powerful concept and establishes a trusted, secure channel between this entity and a peripheral, but it also can be abused.
A participant, \eg the application processor or the secure processor, in control of an attacker, could permanently occupy one or more peripherals resulting in a denial-of-service (DOS) attack.
To mitigate such attacks and manage unresponsive participants, the security monitor has the capability to withdraw access from certain peripherals.
A simple approach to implement this functionality would be to clear the ID field in the peripheral wrapper.
However, this approach is dangerous because it could enable time-of-check to time-of-use (TOCTOU) attacks.
For example, when withdrawing access to a peripheral currently processing a secret, the SM owner would be able to retrieve the result of the request.
To securely withdraw access to certain peripherals, \sanc introduces a withdrawing mechanism, which can be triggered by any entity in the system.
While a withdraw request issued by the SM owner is always granted, a request from other parties first needs to be approved by the security monitor.
When a withdraw request is retrieved by the security monitor, the SM simultaneously starts a timer with a predefined timeout and notifies the owner of the peripheral.
The notification of the peripheral owner is realized by introducing a dedicated interrupt line and a interrupt service routine (ISR) provided by the processors for each peripheral.
The ISR implements a cleanup function responsible for clearing secrets, stopping current transactions, and gracefully releasing the corresponding peripheral.
After the timeout is reached in the security monitor and the peripheral is not released gracefully by the ISR, the SM forcefully releases the peripheral by clearing the ID field in the firewall.

\paragraph{Security Monitor Ownership Transfer.}
A significant advantage of \sanc, compared to other TEE architectures, is the possibility to utilize the TEE infrastructure for various use cases.
While in TEE systems, like ARM TrustZone, one entity, \eg the secure-world, is the exclusive owner of the highest privilege level, \sanc introduces a dynamic ownership transfer mechanism.
The security monitor privilege, which allows the SM owner to configure access privileges and release arbitrary peripherals, can be transferred to any other entity by the SM owner.
To initiate a SM ownership transfer, the entity sends a request with the identifier of the new owner to the security monitor.
The security monitor acknowledges this request by setting the received identifier into the SM owner ID field.
To obtain a clear state, we recommend that the security monitor owner first releases all peripherals.
\subsection{\cpu Design}

Most ARM TrustZone based systems either utilize a dedicated TEE OS to spawn multiple trustlets or exclusively reserve the whole secure domain for one trustlet providing services~\cite{DBLP:journals/csur/PintoS19}.
While running multiple trustlets is tempting, the concept of deploying a fairly complex OS also increases the TCB and the attack surface.
However, as reserving the whole TEE environment for a single trustlet might be lavish, we propose a TEE design in between of these two approaches.
In \cpu, we reduce the TCB to a minimum by providing basic OS services, such as context switches and resource management, directly in hardware.
Although these hardware services cannot provide the same functionality as a rich OS, we argue, that for most TEE use cases, this approach is sufficient.
Typically, trustlets deployed in TEEs offer services with limited complexity, such as a key store, a credential manager, or a cryptographic library~\cite{mandt2016demystifying,johnson2018titan}.
Furthermore this approach also allows us to deeply integrate security features, such as a control-flow integrity~(CFI) unit combined with the secure I/O concept of \sanc, into the processor.
\subsubsection{Control-Flow Integrity}
To protect a program from attacks targeting to alter the control-flow, CFI schemes are commonly used~\cite{DBLP:conf/uss/ZhangS13,DBLP:conf/uss/TiceRCCELP14,DBLP:conf/sp/WangJ10}.
These schemes mitigate attacks like ROP~\cite{DBLP:conf/ccs/Shacham07} or JOP~\cite{DBLP:conf/ccs/CheckowayDDSSW10} by ensuring that the control-flow of the program can not escape the control-flow graph (CFG) determined at compile time.
Enforcing the integrity of the instruction stream can be realized in different degrees of fineness.
While some schemes~\cite{DBLP:journals/tissec/AbadiBEL09,DBLP:journals/tvlsi/AroraRRJ06,DBLP:conf/codaspy/ChristoulakisCA16} preserve the correctness of the execution flow at basic block level, other techniques~\cite{DBLP:conf/cardis/WernerWM15,DBLP:journals/compsec/ClercqGUMV17,DBLP:conf/eurosp/WernerUSM18} maintain the integrity of the control-flow even at instruction granularity.

\paragraph{Control-Flow Integrity Unit.}
\cpu utilizes the existing fine-grain Sponge-Based Control-Flow Protection~(SCFP)~\cite{DBLP:conf/eurosp/WernerUSM18} scheme to protect trustlets within the secure processor from attack attempts.
The main idea of SCFP is to encrypt a program using a sponge-based authenticated encryption primitive during compile time and decrypt it instruction for instruction at runtime.
This method allows SCFP to enforce the, at compile time extracted, CFG at runtime.
Decryption of the individual instructions is realized using a dedicated decryption stage in the processor pipeline.
To successfully decrypt an instruction, the pipeline stage needs to know the key and an internal state.
The SCFP scheme accumulates information over all previously executed instructions in this state.
If the integrity of the state is violated, the decryption fails and returns a faulty instruction, which can be detected with a certain probability by the CPU.
An attacker modifying instructions, \eg using fault injection, or inserting additional instructions, alters the state and can be detected by SCFP.

\subsubsection{Hardware Scheduling}
We extend the native SCFP implementation, which allows to execute a bare-metal program on a processor, to support multitasking.
One approach to enable multitasking with CFI protected trustlets could be realized purely in software using an OS.
However, since TEE operating systems, such as OP-TEE~\cite{OpTee2020}, do not provide state-of-the-art protection mechanisms, such as ASLR or guard pages, mounting an operating system would also increase the attack surface of the TEE.
Therefore, similar to Antikernel~\cite{DBLP:conf/ches/ZonenbergY16}, \cpu offers hardware features to run multiple trustlets on the processor and reduce the software TCB to a minimum.

\paragraph{Hardware Scheduling Unit.}
\cpu introduces a hardware entity providing minimal OS functionality for trustlets.
This hardware unit is responsible for performing secure context switches between individual trustlets in hardware.
The round-robin based scheduling mechanism internally maintains a list of trustlets and after a certain amount of cycles executed, a context switch is conducted.
When performing a context switch, the hardware entity stops the current trustlet, stores the architectural state to a secure place, and loads the next architectural state of the next trustlet.
Additionally, the hardware context switch mechanism also exchanges the decryption key used for SCFP.
Using an individual key for each trustlet yields two major advantages.
First, since the programs are encrypted with a different key, only the developer with the correct key can access the plain program, leading to an IP protection mechanism for trustlets.
Second, using different keys for trustlets enables strong isolation between the applications.
After the context switch, the execution is resumed and the next trustlet is executed.
While this hardware scheduling unit allows the processor to basically consist of several virtual processor cores, only a fixed number of trustlets can run simultaneously on the physical core.
However, since most TEEs are anyway limited in their processing power and only a well-chosen set of trustlets is usually deployed in TEEs by the vendor, an upper bound of processes is acceptable.
Furthermore, we argue that for simple services typically used in mobile platforms, such as a process handling biometric data for unlocking the device or a process handling the secure key storage, no dedicated operating system is needed.
Completely omitting the operating system and providing operating system services using a tiny hardware unit reduces the software TCB to a minimum and would even allow to formally verify the simple hardware unit.
\subsubsection{Control-Flow Integrity with Secure I/O}
The fine granular control-flow integrity unit embedded into the \sanc architecture prevents an attacker from performing control-flow hijack attacks by limiting the control-flow of the program to only valid paths through its control-flow graph.
However, while the CFI scheme protects the control-flow by detecting integrity violation of forward- and backward-edges, and thus prevents attacks such as ROP or JOP, data-oriented attacks can not be detected by this countermeasure.
In such attacks, an adversary modifies control- or non-control related data to break the security of the system.
By manipulating control-related data, such as the condition value in an \texttt{if} statement, the attacker indirectly can influence the control-flow of the program.
Furthermore, an attacker could leak sensitive data, such as passwords or keys, by manipulating addresses in the system.
For example, instead of returning the ciphertext over an API to the user, an attacker could modify the address from pointing to the location of the ciphertext in memory to the encryption key stored in a secure storage element instead by exploiting a buffer overflow.
To lower the impact of such attacks, \cpu binds access to certain assets to a certain CFI state.
More concretely, only when the CFI protected program reaches a predefined CFI state, the program is permitted to access the distinct peripheral.
In \cpu, this strategy is realized by tunneling each interaction request with a peripheral through a dedicated function with a certain state.
\cpu automatically sets the peripheral identifier of the processor to the current state.
Only if this state matches the ID stored in the peripheral wrapper, access to the device is granted.
If an attacker calls the peripheral access function outside the valid control-flow graph, the CFI mechanism detects this violation.
Additionally, if the adversary crafts an address accessing the asset, the state used as an ID does not match the identifier of the peripheral and access to the device is restricted.
\section{Implementation}
\label{sec:implementation}
In this section, we provide a prototype implementation of our \sanc architecture consisting of two cores with two different ISAs.
We first introduce the \riscv lowRISC chip, which we use as our base platform.
Then, we extend this platform with the \sanc features.
Finally, we present our \cpu design and demonstrate how this processor is embedded into the SoC infrastructure.
\subsection{Base Platform}
The prototype implementation is based on the open-source \\lowRISC~\cite{LowRISC2019} project.
Internally, the lowRISC chip consists of the 64-bit \riscv Rocket chip~\cite{NON:Waterman:EECS-2011-62,Asanovic2016} using the RV64GC ISA.
The SoC, capable of running Linux, provides an external off-chip DDR3 memory and an on-chip BRAM.

\subsection{\sanc}
\begin{figure*}[t]
  \begin{center}	
    \includegraphics[width=0.8\linewidth]{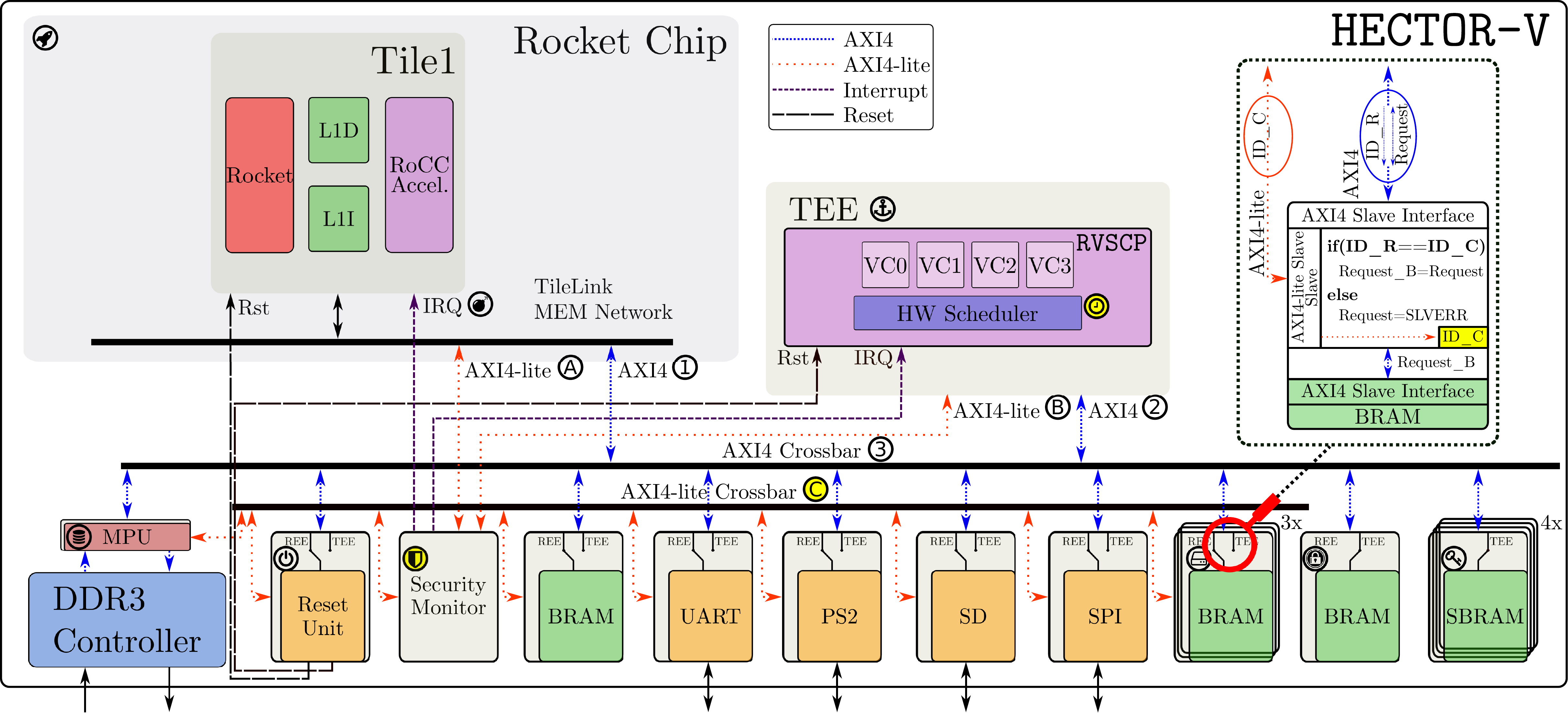}
    \caption{The \sanc architecture.}
    \label{fig:implementation}
  \end{center}
\end{figure*}

Figure~\ref{fig:implementation} depicts the prototype implementation of \sanc.
The extended lowRISC base platform consists of the Rocket Core application processor~\circleds{\faRocket} and the secure processor \cpu integrated into the TEE~\circleds{\faAnchor}.
By providing an AXI4 master interface~\circleds{1}~\circleds{2}, REE and TEE gain access to various peripherals over the shared communication fabric~\circleds{3}.
To differentiate between REE and TEE, the immutable core identifier is directly embedded into this AXI4 master interface.
We further introduce a security monitor~\circleds{\faShield}, a memory protection unit~(MPU)~\circleds{\faDatabase}, a reset unit~\circleds{\faPowerOff}, and a secure storage element~\circleds{\faKey} as part of \sanc.

\paragraph{Interconnect.}
The SoC communication infrastructure consists of a shared AXI4~\circleds{3} and AXI4-lite~\circleds{C} interconnect.
In our architecture, the AXI4 bus is used to enable interaction between the processing cores and the peripherals.
We extended the AXI4 bus protocol and the crossbar to transport the 16-bit user signal along with each bus request.
This user signal carries the 1-bit core ID, the 4-bit process ID, and the 10-bit peripheral ID.
While communication between peripherals and cores requires a high-speed link, the configuration of the security monitor and the peripheral wrappers only consists of small configuration commands.
Hence, we use a lightweight AXI4-lite bus as a configuration channel.
The configuration of the security monitor is realized by introducing a point-to-point communication channel between REE and SM~\circleds{A} and TEE and SM~\circleds{B}.
By using a separate AXI4-lite interconnect~\circleds{C}, the configuration of the peripheral wrappers can only be initiated by the security monitor.
This strategy ensures that neither the TEE nor the REE directly can manipulate the peripheral firewalls; configuration only can be requested over the security monitor module.

\paragraph{Security Monitor.}
To receive commands from REE and TEE, the security monitor implements two AXI4-lite slave interfaces.
The protocol used to interact with the security monitor consists of two privileged and four unprivileged commands.
With the privileged configuration command, the SM owner configures the table entries of the hardware module.
The table includes all peripherals known to the SM, the current claiming status, and a list of identifiers allowed to request access to the device.
By using the privileged ownership transfer configuration command, the SM owner can define a new designated SM owner.
The permission to issue a privileged configuration command is checked by the security monitor using the SM owner ID field stored in the SM module.
The unprivileged commands consist of a claiming and release request command allowing the issuer to gain access to a peripheral.
A status command can be used to determine the permission level of the peripheral and if it is currently claimed.
To gain access to an already claimed peripheral, the unprivileged withdraw request command can be used.
While a withdraw request issued by the SM owner is always granted, the request of the unprivileged participant first needs to be approved.
When the withdraw request is granted, the SM uses dedicated interrupt lines~\circleds{\faBomb} to notify the owner of the peripheral about the withdraw request. 
For each peripheral, a dedicated interrupt line between SM and TEE or REE is introduced.
When the request is approved by the SM, the configuration command is forwarded to the peripheral wrapper over the AXI4-lite crossbar~\circleds{C}.

\paragraph{Peripheral Wrapper.}
An AXI4 read or write request issued by the TEE or REE and transported over the interconnect~\circleds{3} is not directly sent to the peripheral driver.
First, a simple logic deployed in the peripheral wrapper checks if the ID stored in the user signal of the AXI4 request matches the identifier stored in the ID field of the wrapper module.
When the process ID or the peripheral ID is set to zero, the access control is only conducted with the core ID, allowing all entities on the TEE or REE to access the peripheral.
Then, if the ID transported in the request matches the ID stored in the module, the request is forwarded to the actual peripheral.
However, on an ID mismatch, the peripheral wrapper transports the error code \texttt{SLVERR} to the issuer using the \texttt{RRESP} or \texttt{BRESP} AXI4 signal.
In addition to the AXI4 slave interface, the peripheral wrappers also implement an AXI4-lite slave interface.
This interface is used by the SM to set or unset the I in the ID field of the wrapper.

\paragraph{Interrupt.}
The peripheral wrapper also is responsible for routing the interrupt line of the peripheral to the current peripheral owner.
To realize correct interrupt handling, the peripheral wrapper modules consists of one dedicated interrupt line for each participant (REE and TEE).
Based on the ID of the current peripheral owner, the interrupt either is routed to the TEE or the REE.

\paragraph{Software Support.}
To interact with the security monitor, we provide a Linux kernel module for the application processor.
This kernel module allows the user processes to claim, release, withdraw, or query the status of a peripheral.
Furthermore, the kernel module also provides functionalities to configure the security monitor.
To handle interrupts from the security monitor withdraw mechanism, each peripheral driver is extended with a dedicated cleanup interrupt service routine.
This ISR is responsible for clearing any secrets, aborting communication channels with other parties, and releasing the peripheral gracefully using the release mechanism.
For the TEE, we provide a small library.
Similar to the kernel module, this library provides basic functionalities to interact with the security monitor

\paragraph{Physical Memory Protection.}
Isolating a peripheral by binding it exclusively to one entity does not work for the shared, external DDR3 memory.
Therefore, the \sanc architecture introduces a memory protection unit~(MPU)~\circleds{\faDatabase}, which is placed directly between the memory controller and the AXI4 bus interface.
The MPU can be claimed like any other peripheral by the TEE or REE using the security monitor and the dedicated AXI4-lite slave interface.
The party currently claiming the MPU is now able to divide the physical memory into up to 16 regions.
These regions can then be either exclusively accessed by one entity or are shared among multiple entities.
Each incoming AXI4 read or write access is checked by the memory protection unit using the identifier transported with the request.
With this mechanism, a secure storage place for the REE and each virtual core of the \cpu is enabled and shared communication channels between REE and TEE can be established.

\paragraph{Reset Unit.}
The reset unit~\circleds{\faPowerOff} controls the reset lines of the application and the secure processor.
Similar to other peripherals, this entity can be claimed by each participant in the system.
The owner of the device can release the reset lines and turn on or off the other entity in the system.

\paragraph{Secure Storage.}
To securely store sensitive data, such as cryptographic keys, biometric data, or user passwords, \sanc introduces a secure storage element~\circleds{\faKey}.
In contrast to other peripherals, a predefined, immutable identifier consisting of the core ID and the process ID is programmed into the ID field of the wrapper module at design time.
Therefore, only the entity with the corresponding ID can access the storage element.
In the prototype, each of the four virtual cores of \cpu posses an own secure storage.

\subsection{\cpu}
The \cpu prototype implementation is based on the 32-bit \riscv \texttt{REMUS} core~\cite{DBLP:conf/eurosp/WernerUSM18,schaffenrath} with the RV32IMXIE ISA already offering the sponge-based CFI unit.
The \texttt{REMUS} core originally is based on the \texttt{Ri5CY} core~\cite{traber2016pulpino}, which achieves similar performance than a ARM Cortex-M4 core.
We further extend the core with the \cpu features and embed the processor into the \sanc architecture.

\paragraph{TEE Infrastructure.}
By using the AXI4 master interface~\circleds{2} connected to the AXI4 crossbar~\circleds{3}, the \cpu is able to interact with the peripherals, such as the UART or PS2 controller.
Similar to the main application processor, the secure processor implements an AXI4-lite master interface~\circleds{B} to configure and transmit peripheral access requests to the security monitor.

\paragraph{Context Switch.}
The hardware scheduler unit is responsible for performing secure context switches.
For the \cpu prototype, the hardware scheduler maintains a list of four slots enabling four virtual cores \texttt{VC0} \dots \texttt{VC3} on the \cpu core.
On a context switch, the hardware scheduler saves the current SCFP state and the current register file and loads the state, the decryption key, and the register file for the next trustlet.
To implement the replacement of the register file, we added four additional register sets to our processor.
Note that the register file replacement could also be implemented by storing and loading the content of the registers to memory, \eg to a secure storage element, to keep the area overhead of the processor low.
To differentiate between the four trustlets executed on the \cpu, each of the four slots gets assigned an individual process ID.
While the core ID is identical for all slots, the hardware scheduler replaces the process ID directly in the AXI4 and AXI4-lite master interface individually for each slot.
By using the same core ID for all four threads, a peripheral could be configured to be accessible by all four trustlets.

\paragraph{Decryption Keys.}
To decrypt the encrypted instruction stream, the SCFP unit needs to know the decryption key for the corresponding trustlet.
In the prototype implementation, the key is stored in a dedicated control and status register of each virtual core, which is only accessible from the respective core.
To initially load the key into the secure storage, our prototype trustlet consists of a small, unencrypted boot code and the actual, encrypted code.
The unencrypted boot code can either generate the key, load the key from the network, or directly from the binary.
After storing the key into the key register, the actual encrypted trustlet starts to execute.

\paragraph{Code Storage.}
Each of the virtual processor cores \texttt{VC0} \dots \texttt{VC3} running on the \cpu processor executes code from an on-chip BRAM.
While \texttt{VC1} \dots \texttt{VC3} fetch code from a claimable BRAM~\circleds{\faHddO}, the first virtual processor core fetches its code from a secure code storage element~\circleds{\faExpeditedssl} exclusively and permanently owned by \texttt{VC0}.
To utilize one of the virtual cores at \cpu as an enclave, the issuer needs to store the trustlet code on the BRAM of this core.
Since the code of the first virtual core is stored in a secure storage element~\circleds{\faExpeditedssl}, this code cannot be changed by the REE or by the other virtual cores of \cpu at any point in time.

\paragraph{Control-Flow Integrity with Secure I/O.}
To implement the concept of binding access to peripherals to a certain CFI state, each AXI4 request leaving the \cpu is annotated with the current CFI state.
The processor directly places a compressed form of the current state into the 10-bit peripheral ID field of the AXI4 user signal.
\begin{figure}[t]
  \begin{center}	
    \includegraphics[width=0.7\linewidth]{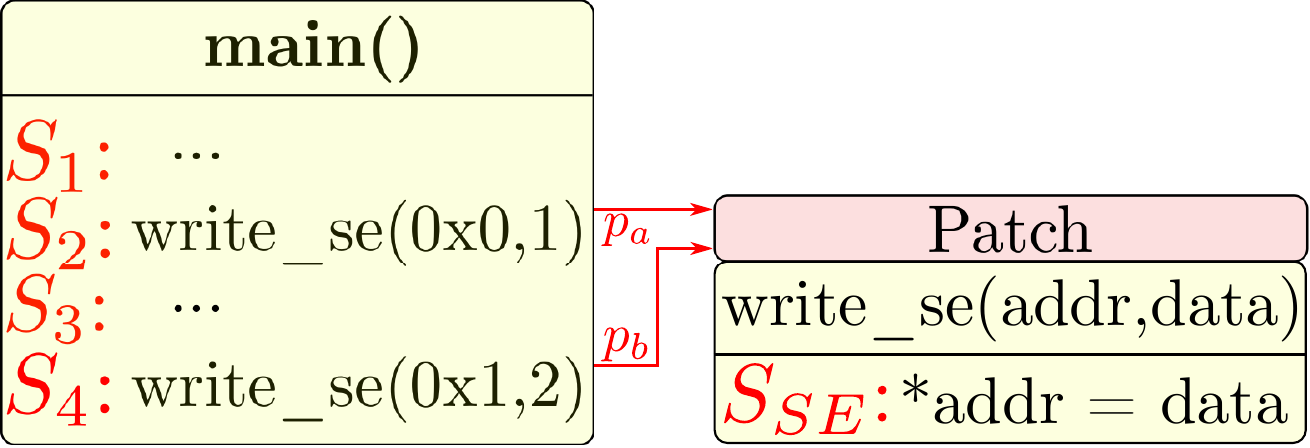}
    \caption{Access function for secure peripheral accesses.}
    \label{fig:access_function}
  \end{center}
\end{figure}
To only allow, \eg write access to the secure element when reaching a predefined state, the trustlet tunnels all write accesses to this peripheral through the function \texttt{write\_se}.
Since decrypting this function only works, when a certain state $S_{SE}$ is reached, the SCFP scheme automatically patches the state of the callee with the patch value $p_a$ or $p_b$.
As seen in Figure~\ref{fig:access_function}, the patching mechanism of SCFP ensures that, on each valid call of the access function, state $S_{SE}$ is reached.
By using this state as the peripheral identifier, access to a specific peripheral only is granted when reaching state $S_{SE}$ through the access function.
Similar to the setup procedure of the decryption keys, the boot code of the trustlet claims the peripheral used in the code by sending the identifier consisting of the core identifier (ID of \cpu), the process identifier (ID of the virtual processor), and the peripheral identifier (compressed CFI state $S_{SE}$) to the security monitor.
Now, access to secrets stored in the secure storage element or other sensitive peripherals, like a fingerprint reader, only is permitted when reaching a predefined CFI state.
If a trustlet does not need explicit protection of certain peripherals, the peripheral ID in the peripheral claiming request is set to zero.

\subsection{Performance Evaluation}
In this section, we analyze the latency for peripheral and TEE communication, as well as the runtime overhead of the TEE.

\paragraph{Peripheral Access Latency.}
The secure I/O design of \sanc requires the entities to claim resources, \eg the BRAM, before accessing them.
However, this claiming process, which is initialized in software, only needs to be conducted once for each entity and resource.
To determine the performance of this mechanism, we configured a trustlet on the \cpu to be the designated security monitor~(SM) owner.
We measured an execution time of 188 cycles on average for sending a claim request from the REE to the SM and reading back this status.
Furthermore, the secure I/O concept of \sanc introduces the peripheral wrappers, which compare the identifier of each incoming AXI4 request to the identifier stored in the register of this wrapper.
To measure the performance impact of this filtering mechanism, we copied data from the REE to a protected BRAM.
However, since the ID comparison only requires a simple hardware logic, we could not measure any performance impact when comparing to a platform without the peripheral wrappers.

\paragraph{Communication Latency between REE and TEE.}
The current implementation of \sanc utilizes a memory region shared between REE and the trustlet for communication.
Hence, the communication latency between these two security domains is determined by the performance of the external DDR3 memory and the communication API arranged by the trustlet and the REE application.
To characterize the general performance of the underlying memory subsystem, we benchmarked the external memory using LMBench~\cite{DBLP:conf/usenix/McVoyS96}.
We measured an average memory access latency of \SI{876}{\nano\second} for the \texttt{lat\_mem\_rd 64M 512} testcase.
To measure the communication latency, \ie the timeframe of sending a message from REE to TEE and receiving the acknowledgment on the REE, we installed a benchmarking trustlet on \cpu.
We configured this trustled to establish a shared memory region on the external memory using the secure I/O design of \sanc.
This trustlet acknowledges the received \SI{1}{\mega\byte} message by writing a status value in the shared memory.
On the REE, we implemented a blocking function allowing to send data to the shared memory region and waiting for the acknowledgment from the trustlet.
Note that the usage of blocking functions for communication also is recommended by the popular GlobalPlatform TEE API specification~\cite{GlobalPlatformTEE} used by many TEE operating systems, such as OP-TEE~\cite{OpTee2020}.
We measured an execution time of $5498$ cycles on average for transmitting a \SI{1}{\mega\byte} memory object to the TEE and receiving the response from the REE.
For comparison, simply writing a \SI{1}{\mega\byte} to the external memory already takes $5261$ cycles on average.
This timing difference of 4.5\,\% on average is negligible, as the memory performance is the major performance indicator for cross-domain communication.
Although the usage of blocking functions is common for TEEs, \sanc could further be extended to implement an interrupt-based mailbox system similar to Apple's SEP~\cite{mandt2016demystifying}.

\paragraph{\cpu Runtime Overhead.}
For the secure \cpu processor, we extended the existing core implementing SCFP~\cite{DBLP:conf/eurosp/WernerUSM18} with the hardware scheduler responsible for scheduling the trustlets and managing the identifiers.
As these changes do not alter the single thread performance of the core, we observed same performance numbers for \cpu.
\citeauthor{DBLP:conf/eurosp/WernerUSM18} measured a performance overhead of 9.1\,\% on average for protecting a broad range of macrobenchmarks with CFI.
The baseline for these measurements is the \texttt{Ri5CY}~\cite{traber2016pulpino} core, which achieves similar performance than a state-of-the-art ARM Cortex-M4~\cite{schiavone2017slow}.
Hence, \cpu is suitable for executing a vast variety of trustlets.

\subsection{Area Overhead}
To quantify the area overhead introduced by the additional \riscv core and the \sanc features, we synthesized the lowRISC base platform for a Xilinx Kintex-7 series FPGA.
More concretely we synthesized the extended lowRISC platform with the additional \sanc features and a selection of different \riscv cores used as TEE processor.
\begin{table}[b]
\centering
\caption{Lookup table~(LUT) overhead of the overall architecture consisting of the \sanc features and an additional secure processor compared to the lowRISC base project.}
\label{tab:areaoverheadtotal}
\begin{tabular}{lll}
\hline
\textbf{Configuration}            & \textbf{\begin{tabular}[c]{@{}l@{}}Area\\ {[}LUTs{]}\end{tabular}} & \textbf{\begin{tabular}[c]{@{}l@{}}Area Overhead\\ {[}\%{]}\end{tabular}} \\ \hline
lowRISC base platform                     & 55,443                                                             & -                                                                         \\
\sanc with \texttt{RI5CY}        & 63,648                                                             & 14.8                                                                      \\
\sanc with  \texttt{REMUS}        & 67,024                                                                  & 20.89                                                                         \\
\sanc with  \texttt{FRANKENSTEIN} & 68,746                                                             & 23.99                                                                     \\ \hline
\end{tabular}
\end{table}
Table~\ref{tab:areaoverheadtotal} shows the total hardware overhead of the \sanc architecture consisting of either the \texttt{Ri5CY}~\cite{traber2016pulpino}, the \texttt{REMUS}~\cite{DBLP:conf/eurosp/WernerUSM18,schaffenrath}, or the \texttt{FRANKENSTEIN}~\cite{DBLP:conf/acsac/SchillingWNM18} core used as secure TEE processor.
The \texttt{REMUS} core, which is based on the \texttt{Ri5CY}, implements a decryption unit to offer the Sponge-Based Control-Flow Protection~(SCFP) mechanism.
\texttt{FRANKENSTEIN}, which is an extended version of the \texttt{REMUS} core, is a 64-bit \riscv processor including the SCFP scheme and a pointer protection scheme for secure memory accesses.
While the Rocket Core used in the lowRISC platform is capable of running Linux, the processor core still is rather small compared to application processors from vendors like Intel, ARM, and AMD.
Therefore, when using a larger processor in the \sanc architecture, the relative overhead introduced by the additional secure TEE processor is negligible.
\begin{table}[b]
\centering
\caption{Hardware overhead of different features.}
\label{tab:lutsfeatures}
\begin{tabular}{lll}
\hline
\textbf{Component} & \textbf{\begin{tabular}[c]{@{}l@{}}Area\\ {[}LUTs{]}\end{tabular}} & \textbf{\begin{tabular}[c]{@{}l@{}}Area\\ {[}\%{]}\end{tabular}} \\ \hline
Rocket Chip        & 33,341                                                             & 52.38                                                            \\
\texttt{RI5CY}              & 5,780                                                              & 9.08                                                             \\
Security Monitor   & 446                                                                & 0.7                                                              \\
Peripheral Wrapper & 43                                                                 & 0.07                                                             \\
AXI4 Crossbar      & 3,052                                                              & 4.79                                                             \\
AXI4-lite Crossbar & 93                                                                 & 0.14                                                             \\ \hline
\end{tabular}
\end{table}
In Table~\ref{tab:lutsfeatures}, we list area number for different components in the \sanc architecture using the \texttt{RI5CY} core as secure TEE processor.

\section{Use Cases}
\label{sec:use_case}

This section proposes use cases for the secure TEE design consisting of the \sanc architecture and the \cpu.
More concretely, we demonstrate how the TEE can be used to boot Linux on the main application processor securely.
After the secure boot use case, we show how the \cpu can be utilized to securely execute trustlets.
Note that these two scenarios are only a selection of many other use cases that can be realized with \sanc.
\subsection{Secure Boot}
On the unmodified lowRISC platform, a zero stage bootloader~(ZSBL) permanently stored in the on-chip BRAM first loads the Berkeley boot loader~(BBL) from the external SD-card.
Then, the ZSBL hands over control to the BBL first stage bootloader~(FSBL).
The BBL, which is linked against the Linux kernel, fetches the Linux image and sets up the environment by configuring the hardware threads~(HARTS) and the memory.
Finally, the Linux operating system starts.
However, loading the FSBL and the Linux image directly from the SD-card to boot up the device is problematic in many ways.
An attacker with physical access to the device can boot arbitrary code by simply modifying the unprotected boot files stored on the SD-card.
This attack methodology also can be used in an online attack by overwriting the boot files.
To only allow authenticated software to boot, vendors frequently embed a secure boot mechanism into their systems.
Here, a chain-of-trust is generated by authenticating each boot file before executing it.

In our use case implementing secure boot using features of the \sanc architecture, the first virtual core \texttt{VC0} of \cpu is the designated security monitor owner.
At design time, the reset unit of the system is configured to release the reset line of \cpu and keep the REE processor halted when applying power to the SoC.
Additionally, the security monitor owned by \texttt{VC0} is unconfigured, except for the reset unit which is claimed by the SM owner.
When applying power to the device, \texttt{VC0} starts to execute the ZSBL stored in the secure code storage~\circleds{\faExpeditedssl}.
Since the secure code storage is permanently and exclusively owned by \texttt{VC0}, the ZSBL is isolated from the other parties and only \texttt{VC0} can update this code using an update mechanism.
Due to these strong protection mechanisms, the ZSBL stored in the secure code storage is the system root-of-trust.
The ZSBL code first claims the SD-card controller and configures the memory protection unit of the DDR3 memory.
Then, the ZSBL fetches the BBL from the SD card and stores it to the external memory.
Before passing control to the BBL, the ZSBL determines the hash value of the loaded BBL image.
When this hash value does not match the expected hash value stored in the secure storage element~\circleds{\faKey} of \texttt{VC0}, the boot process is aborted.
Again, an update of the expected hash value only can be initiated by the first virtual processor core of \cpu because the secure storage element exclusively is owned by this party and can not be claimed by any other party at any point in time.
If the verification of the BBL was successful, the ZSBL releases the SD-card driver, gives the main application processor access to the DDR3 memory region by configuring the MPU, and triggers the reset unit to start the REE.
Now, the Rocket Core starts to execute the modified BBL from the external memory.
The modified BBL then requests access to the SD-card controller, loads the Linux image from the SD-card, and verifies the loaded image by comparing the computed hash value with the expected hash value.
Finally, if the verification process succeeds, the Linux operating system starts and can claim the first peripherals by sending requests to the security monitor.
Since each stage of the boot process now is cryptographically verified, the system is in a genuine state.
Now, the \texttt{VC0} passes the SM owner privilege to the main application processor~(AP).
This change is initiated by sending the identifier of the AP to the SM using the ownership transfer command.
Although the REE is now in full control of the system, \eg is able to switch of the secure processor using the reset unit, secrets stored in the secure storage~\circleds{\faKey} and the secure code storage~\circleds{\faExpeditedssl} are still protected from AP accesses.
\subsection{Trustlets}
In this use case, we utilize the virtual processor \texttt{VC1} of \cpu to execute a trustlet.
We compile the trustlet with a LLVM-based toolchain supporting the SCFP scheme and encrypt the program with the key $K$.
To interact with the outside and to store secrets, the trustlet uses the UART controller and the secure storage element.
The secure storage element is protected from malicious accesses by tunneling all requests through a dedicated access function and binding the access to it to the CFI state $S_{SE}$.
We define access to the UART controller as uncritical.
Therefore the interaction with this controller is unprotected.
First, the application processor switches off the \cpu by utilizing the reset unit.
Then, the processor claims a claimable BRAM~\circleds{\faHddO} and stores the code, consisting of an unencrypted small boot code and the encrypted trustlet, to this storage.
After the code is saved, the AP sets \texttt{VC1} as owner of the code storage in the security monitor.
Furthermore, the MPU is configured to provide a memory region owned by the trustlet to use as RAM and a shared memory region to establish a communication channel between REE and the trustlet.
Finally, the \cpu is started by releasing the reset line of the secure processor and code gets executed.
To initialize key $K$ and state $S_{SE}$, the boot code of the trustlet writes the decryption key into the key register of $VC1$ and claims the secure key storage element by setting the compressed state $S_{SE}$ into the peripheral ID field.
Now, control is passed to the SCFP protected trustlet code and the decryption stage of the processor decrypts the code using the key.
To access the UART controller, the trustlet needs to claim the controller by sending a claim request to the SM.
The secure storage element is accessed by using the dedicated access function.
By setting the compressed state directly into the peripheral ID field of each AXI4 request leaving the processor, access to the secure storage element only is permitted when reaching the predefined state $S_{SE}$.
The communication channel established using the shared, external memory allows the REE and trustlet to exchange information.
When the designated SM owner, the AP, withdraws access to the UART controller, the claimable BRAM, or the shared memory region, the trustlet is notified using an interrupt.
This interrupt is handled by the ISR implemented on the trustlet.
The ISR clears all secrets, aborts communication with the UART controller, gracefully releases the peripheral, and enters a predefined IDLE state.

\section{Security Discussion}
\label{sec:sec}
In this section, we analyze security guarantees of \sanc and discuss security properties of the secure \cpu processor.
\subsection{\sanc}
The security guarantees of \sanc are established with the introduced architectural features, \ie the security monitor, the peripheral wrappers, and the separation of the REE and the TEE in the SoC.

\paragraph{Isolation between REE and TEE.}
In each TEE design, the key challenge is to guarantee strong isolation between REE and TEE.
Most TEE designs fail to provide this guarantee by either be vulnerable to side-channel and microarchitectural attacks or by insufficiently protecting TEE resources, such as the memory or peripherals~\cite{DBLP:journals/ieeesp/KostiainenDC20}.
\sanc mitigates all cache- and microarchitectural-based side-channel attacks by separating REE and TEE using dedicated cores for each domain.
This approach ensures that no sensitive components, such as caches, branch predictors, and the execution pipelines, are shared between REE and TEE.
To protect resources, \sanc provides architectural features to bind a peripheral to a entity using the SM and the peripheral wrappers.
In \sanc, for each entity, \ie the REE or the trustlets on the TEE, a dedicated memory region can be reserved.
This memory region is isolated using hardware-enforced access checks in the MPU using the ID of the entity.

\paragraph{Trusted Computing Base.}
The TCB of \sanc consists of several hardware and software features, which are marked in yellow in Figure~\ref{fig:implementation}.
The central trust anchor is the security monitor~\circleds{\faShield}, which manages access permissions to peripherals and the memory.
Here, the SM configures the peripheral wrappers over the exclusively owned AXI4-lite~\circleds{C} interconnect by setting the ID of the authorized entity into the ID register.
In \sanc, the SM only can be configured by one entity, the SM owner.
This entity, the software TCB, either is a trustlet in the \cpu or a kernel task in the REE and is responsible for configuring the security monitor.
The SM differentiates between the SM owner and the other parties using the IDs.
While the REE ID is permanently and immutable programmed into the AXI4 interface~\circleds{1} of this core, the identifiers of the trustlets in \cpu are managed by the hardware scheduler~\circleds{\faClockO} TCB.

\paragraph{Malicious Software TCB.}
Although the SM owner is responsible for configuring the SM and is, therefore, part of the software TCB, \sanc considers this entity as potentially malicious and provides architectural features to limit the impact of this threat.
A malicious SM owner either could be the result of an attacker with kernel privileges on the REE currently being the SM owner or a malicious trustlet in the TEE with the SM ownership.
Here, the goal of the attacker is to threaten the confidentiality, integrity, and availability of the system by misconfiguring the SM.
\sanc maintains the integrity and confidentiality of secrets stored in peripherals, \eg the external memory or internal BRAMs, by exclusively binding these elements to one entity.
The hardware-enforced ID access check prevents all other entities, even the SM owner, to access these elements.
Furthermore, to protect critical secrets, such as keys, hash values, or the secure boot code, \sanc offers the possibility to set an ID of an entity immutable into the ID field of the critical element, preventing the SM owner and other parties from accessing this element.
Although the SM owner can initiate a peripheral release request, this hardware mechanism first notifies the current owner of the peripheral about the incoming withdraw procedure, allowing to clear sensitive data.
However, a malicious SM owner could influence the availability of the system by permanently withdrawing peripherals.
To prevent the SM owner to switch off a different entity in the system and then access secrets stored in the peripherals, the reset unit could, similar to the withdrawing mechanism, notify the entity about the incoming reset.
\subsection{\cpu}
The \cpu processor, which is tightly integrated into the \sanc architecture, consists of several hardware features, such as a CFI unit coupled with secure I/O and a hardware scheduler.
\paragraph{Isolation between Trustlets.}
In addition to the isolation guarantees between REE and TEE, the trustlets within the secure processor also need strong isolation between each other.
The hardware scheduler, which acts as TCB for the \cpu, is responsible of scheduling the trustlets.
Here, on each context switch, this unit replaces the identifiers of each trustlets.
As this swapping mechanism is realized in hardware, the trustlets cannot influence the ID in software.
Isolating resources, such as the external memory, code storage, and peripherals, is realized by claiming separate resources for each trsutlet using the identifier-based secure I/O mechanism.
While the heterogeneous architecture prevents an attacker from performing cache and transient-based attacks between REE and TEE, \cpu needs additional countermeasures, such as flushing the microarchitectural state~\cite{DBLP:journals/corr/abs-2005-02193} on a context switch, to also protect against cache attacks.
Due to the simplicity and openness of \cpu, transient-based attacks cannot be performed by an adversary.

\paragraph{Communication Interface.}
To establish a communication between REE and TEE, a shared memory region for the trustlet and the REE application is created using the MPU and the ID of both parties.
Recent attacks~\cite{WoW2016,Bits2016} demonstrated that a single bug, \eg a buffer overflow, in the software interface between REE and TEE could completely compromise the secure domain.
To prevent the exploitation of such vulnerabilities, the trustlets are protected using hardware features of \cpu.
The CFI unit prevents an attacker from performing control-flow hijacking attacks, such as ROP or JOP.
Furthermore, \cpu limits the impact of data-oriented attacks, which cannot be detected by CFI units.
Here, \cpu protects trustlets and resources from these attacks by restricting access to peripherals only when reaching a predefined CFI state.
Moreover, the attack surface of \cpu is minimized by completely omitting a TEE OS and using a hardware TCB controlling the context switches and the resource management.
In addition to logical attacks on the interface, the \sanc approach also aggravates physical bus probing and sniffing attacks on the communication interface~\cite{bussniffing} by integrating the TEE into the SoC.

\section{Related Work}
This section summarizes TEEs introduced by both, academia and industry, and compares them to \sanc.

\paragraph{Intel SGX.} 
In Intel Software Guard Extensions~\cite{SGX2019}, an enclave is created purely in software and protected with hardware features, such as a memory encryption engine~(MEE).
Although this approach allows to flexibly spawn new enclaves in software, SGX suffers from several disadvantages.
First, Intel explicitly excludes side-channel and fault attacks in their threat model~\cite{IntelSidechannel} and this weakness already is actively exploited~\cite{DBLP:conf/woot/BrasserMDKCS17,DBLP:conf/dimva/SchwarzWGMM17,DBLP:conf/eurosec/GotzfriedESM17,DBLP:conf/uss/BulckMWGKPSWYS18,van2020lvi, DBLP:journals/ieeesp/ChenCXZLL20,murdock2020plundervolt}.
Second, SGX does not natively protect user I/O and is therefore vulnerable to attacks on peripherals.

\paragraph{ARM TrustZone.} 
In this TEE design, ARM enforces a secure and non-secure domain within the processor using a virtual processor approach based on an identifier, the non-secure~(NS) bit.
The enforcement of the security policy is directly implemented in the AXI interface of the SoC components.
This security policy can be static for some peripherals, like the fingerprint reader, or can be dynamically configured by the secure world using several hardware blocks~\cite{DBLP:journals/csur/PintoS19}.
The TrustZone Protection Controller~(TZPC), which solely can be configured by the secure world, acts as a root-of-trust and manages access permissions for single peripherals~\cite{silva2019arm}.
To partition memory into secure and non-secure areas, the secure world can configure the security policies using the TZ Address Space Controller (TZASC) and the TZ Memory Adapter (TZMA).
Since operating systems deployed in TrustZone typically do not provide state-of-the-art defense strategies~\cite{cerdeira2020sok}, a malicious trustlet could threaten the security of the secure world.
Therefore, vendors integrating TZ into their products limit their systems to only a pre-defined set of trustlets.
SANCTUARY~\cite{DBLP:conf/ndss/BrasserGJSS19} tackles this problem by extending TrustZone to support user-space applications.
Here, SANCTUARY utilizes TZ to exclusively reserve a core and memory to spawn individual Sanctuary instances.
These instances, which allow users to deploy own trustlets, mitigate cache-based side-channel attacks by using the L1 cache of the reserved core and excluding to use the L2 cache.
Although SANCTUARY enhances TrustZone to support user-defined trustlets, it inherits limitations of the underlying architecture.
In contrast to \sanc, TrustZone only uses a 1-bit ID to differentiate between secure and non-secure world.
Hence, TrustZone only can bind peripherals to a core or to the secure domain and fails to offer fine-granular protection of peripherals for multiple trustlets~\cite{DBLP:journals/corr/abs-2010-15866}.
Furthermore, TrustZone is an inflexible approach with the secure domain as the static trusted entity.
\sanc allows to dynamically switch the trusted domains between the entities, which enhances potential use cases for the TEE design.
Additionally, \sanc offers hardware features, such as a fine-granular CFI unit combined with secure I/O, to isolate trustlets within the secure domain.
Although SANCTUARY mitigates cache-based side-channel attacks by reserving a core and its L1 cache for a trustlet, other ARM TrustZone based security architecture are still vulnerable against such attacks~\cite{DBLP:conf/uss/LippGSMM16}.

\paragraph{RISC-V based TEEs.}
The introduction of several open-source RISC-V cores mobilizes research on open TEE solutions.
SANCTUM~\cite{DBLP:conf/uss/CostanLD16} aims to offer a similar programming model to SGX on RISC-V.
This design uses a software-based security monitor and requires minimal hardware changes to dynamically spawn new enclaves.
In addition to SGX, SANCTUM mitigates cache-based side-channel attacks by using cache partitioning for each enclave.
Similarly, Keystone~\cite{DBLP:conf/eurosys/LeeKSAS20} utilizes a security monitor to enforce TEE guarantees and uses the physical memory protection~(PMP) feature of RISC-V to isolate individual enclaves.
To mitigate software side-channel attacks, Keystone flushes enclave states on a context switch.
Although these designs address the side-channel problematic of TrustZone and SGX, SANCTUM and Keystone do not provide architectural features for secure I/O, leaving communication with peripherals unprotected.

\paragraph{SiFive WorldGuard.}
Concurrent developed with the HECTOR-V architecture, SiFive recently introduced WorldGuard~\cite{WorldGuard19}.
Here, each core gets assigned a world ID and each process on the core can be annotated with a process ID.
This ID then is transported using the interconnect and requests from participants are filtered by peripherals, the memory, and the caches.
Similar to \sanc, this approach allows WorldGuard to reserve one core exclusively for the secure domain to provide strong isolation guarantees.
However, both architectures differ in various design choices.
First, \sanc uses a hardware-based security monitor which only can be configured by one party.
In contrast to WorldGuard, the security monitor ownership can be dynamically transferred to other parties allowing flexible use cases.
Additionally, the security monitor allows each participant in \sanc to request access to certain peripherals and request access to already claimed peripherals using a withdraw request.
We further propose a concrete secure processor design utilizing features of the heterogeneous architecture to create a trusted execution environment.
Moreover, we comprehensively describe the hardware-software interaction and demonstrate features of \sanc by introducing several use case scenarios.
\paragraph{GPU based TEEs.}
In addition to the proprietary solutions from Apple, Microsoft, and SiFive, several academic TEEs based on GPUs have been introduced recently.
HIX~\cite{DBLP:conf/asplos/JangTKSH19} and Graviton~\cite{DBLP:conf/osdi/VolosVB18} utilize the GPU to establish a trusted execution environment for trustlets.
Although these designs share the idea of \sanc to use a heterogeneous system for the TEE, these GPU-based architectures do not support secure I/O for peripherals.
\section{Conclusion}
\label{sec:conclusion}
In this paper, we proposed \sanc, a secure TEE design strategy using a heterogeneous CPU architecture.
Our design establishes secure paths between peripherals and the cores by tagging each party with an identifier.
The peripherals enforce access permissions by checking the ID, which is transported along with each bus request.
To configure these access permissions, we integrate a hardware-based security monitor into the architecture.
The security monitor, which exclusively can set permissions, is owned by a configuration party.
By allowing to transfer this ownership to other parties, \sanc allows flexible permission management.
In contrast to similar design approaches, we provide a notifier-based mechanism to withdraw access to certain peripherals securely.
We further introduce \cpu, a security-hardened CPU design tailored for our architecture.
\cpu combines a fine-granular control-flow integrity scheme with the secure I/O concept of \sanc to restrict access to assets.
To complete our TEE design, we introduce secure data and code storage elements, a reset unit, and a memory protection unit.
We examine the features of our architecture in a secure boot and enclave scenario.

\begin{acks}
This project has received funding from the European Research Council (ERC) under the European Union’s Horizon 2020 research and innovation programme (grant agreement No 681402) and by the Austrian Research Promotion Agency (FFG) via the competence center Know-Center (grant number 844595), which is funded in the context of COMET - Competence Centers for Excellent Technologies by BMVIT, BMWFW, and Styria.
\end{acks}

\bibliographystyle{ACM-Reference-Format}
\balance
\bibliography{main}


\begin{thebibliography}{66}


\ifx \showCODEN    \undefined \def \showCODEN     #1{\unskip}     \fi
\ifx \showDOI      \undefined \def \showDOI       #1{#1}\fi
\ifx \showISBNx    \undefined \def \showISBNx     #1{\unskip}     \fi
\ifx \showISBNxiii \undefined \def \showISBNxiii  #1{\unskip}     \fi
\ifx \showISSN     \undefined \def \showISSN      #1{\unskip}     \fi
\ifx \showLCCN     \undefined \def \showLCCN      #1{\unskip}     \fi
\ifx \shownote     \undefined \def \shownote      #1{#1}          \fi
\ifx \showarticletitle \undefined \def \showarticletitle #1{#1}   \fi
\ifx \showURL      \undefined \def \showURL       {\relax}        \fi
\providecommand\bibfield[2]{#2}
\providecommand\bibinfo[2]{#2}
\providecommand\natexlab[1]{#1}
\providecommand\showeprint[2][]{arXiv:#2}

\bibitem[\protect\citeauthoryear{??}{CVE}{2018}]%
        {CVESpi}
 \bibinfo{year}{2018}\natexlab{}.
\newblock
  \bibinfo{title}{\href{http://cve.mitre.org/cgi-bin/cvename.cgi?name=CVE-2016-10423}{{CVE}-2016-10423.}}
\newblock
\newblock


\bibitem[\protect\citeauthoryear{Abadi, Budiu, Erlingsson, and Ligatti}{Abadi
  et~al\mbox{.}}{2009}]%
        {DBLP:journals/tissec/AbadiBEL09}
\bibfield{author}{\bibinfo{person}{Mart{\'{i}}n Abadi}, \bibinfo{person}{Mihai
  Budiu}, \bibinfo{person}{{\'{U}}lfar Erlingsson}, {and} \bibinfo{person}{Jay
  Ligatti}.} \bibinfo{year}{2009}\natexlab{}.
\newblock
  \showarticletitle{\href{https://doi.org/10.1145/1609956.1609960}{Control-flow
  integrity principles, implementations, and applications}}.
\newblock \bibinfo{journal}{\emph{{ACM} Trans. Inf. Syst. Secur.}}
  \bibinfo{volume}{13} (\bibinfo{year}{2009}).
\newblock


\bibitem[\protect\citeauthoryear{Andzakovic}{Andzakovic}{2019}]%
        {bussniffing}
\bibfield{author}{\bibinfo{person}{Denis Andzakovic}.}
  \bibinfo{year}{2019}\natexlab{}.
\newblock
  \bibinfo{booktitle}{\emph{\href{https://pulsesecurity.co.nz/articles/TPM-sniffing}{Extracting
  Bitlocker Keys from a TPM}}}.
\newblock


\bibitem[\protect\citeauthoryear{Apple}{Apple}{2020}]%
        {AppleT2}
\bibfield{author}{\bibinfo{person}{Apple}.} \bibinfo{year}{2020}\natexlab{}.
\newblock
  \bibinfo{booktitle}{\emph{\href{https://support.apple.com/en-us/HT208862}{About
  the Apple T2 Security Chip}}}.
\newblock


\bibitem[\protect\citeauthoryear{ARM}{ARM}{2019}]%
        {axi}
\bibfield{author}{\bibinfo{person}{ARM}.} \bibinfo{year}{2019}\natexlab{}.
\newblock \showarticletitle{AMBA AXI and ACE Protocol Specification}.
\newblock \bibinfo{journal}{\emph{arm.com}} (\bibinfo{year}{2019}).
\newblock


\bibitem[\protect\citeauthoryear{ARM}{ARM}{2020}]%
        {ARMbigLittle}
\bibfield{author}{\bibinfo{person}{ARM}.} \bibinfo{year}{2020}\natexlab{}.
\newblock
  \bibinfo{booktitle}{\emph{\href{https://www.arm.com/why-arm/technologies/big-little}{Processing
  Architecture for Power Efficiency and Performance}}}.
\newblock


\bibitem[\protect\citeauthoryear{ARM}{ARM}{2009}]%
        {arm2009security}
\bibfield{author}{\bibinfo{person}{Architecure ARM}.}
  \bibinfo{year}{2009}\natexlab{}.
\newblock \showarticletitle{Security technology building a secure system using
  trustzone technology (white paper)}.
\newblock \bibinfo{journal}{\emph{ARM Limited}} (\bibinfo{year}{2009}).
\newblock


\bibitem[\protect\citeauthoryear{Arora, Ravi, Raghunathan, and Jha}{Arora
  et~al\mbox{.}}{2006}]%
        {DBLP:journals/tvlsi/AroraRRJ06}
\bibfield{author}{\bibinfo{person}{Divya Arora}, \bibinfo{person}{Srivaths
  Ravi}, \bibinfo{person}{Anand Raghunathan}, {and} \bibinfo{person}{Niraj~K.
  Jha}.} \bibinfo{year}{2006}\natexlab{}.
\newblock
  \showarticletitle{\href{https://doi.org/10.1109/TVLSI.2006.887799}{Hardware-Assisted
  Run-Time Monitoring for Secure Program Execution on Embedded Processors}}.
\newblock \bibinfo{journal}{\emph{{IEEE} Trans. Very Large Scale Integr.
  Syst.}}  \bibinfo{volume}{14} (\bibinfo{year}{2006}).
\newblock


\bibitem[\protect\citeauthoryear{Asanovic, Avizienis, Bachrach, Beamer,
  Biancolin, Celio, Cook, Dabbelt, Hauser, Izraelevitz, Karandikar, Keller,
  Kim, Koenig, Lee, Love, Maas, Magyar, Mao, Moreto, Ou, Patterson, Richards,
  Schmidt, Twigg, Vo, and Waterman}{Asanovic et~al\mbox{.}}{2016}]%
        {Asanovic2016}
\bibfield{author}{\bibinfo{person}{Krste Asanovic}, \bibinfo{person}{Rimas
  Avizienis}, \bibinfo{person}{Jonathan Bachrach}, \bibinfo{person}{Scott
  Beamer}, \bibinfo{person}{David Biancolin}, \bibinfo{person}{Christopher
  Celio}, \bibinfo{person}{Henry Cook}, \bibinfo{person}{Daniel Dabbelt},
  \bibinfo{person}{John Hauser}, \bibinfo{person}{Adam Izraelevitz},
  \bibinfo{person}{Sagar Karandikar}, \bibinfo{person}{Ben Keller},
  \bibinfo{person}{Donggyu Kim}, \bibinfo{person}{John Koenig},
  \bibinfo{person}{Yunsup Lee}, \bibinfo{person}{Eric Love},
  \bibinfo{person}{Martin Maas}, \bibinfo{person}{Albert Magyar},
  \bibinfo{person}{Howard Mao}, \bibinfo{person}{Miquel Moreto},
  \bibinfo{person}{Albert Ou}, \bibinfo{person}{David~A Patterson},
  \bibinfo{person}{Brian Richards}, \bibinfo{person}{Colin Schmidt},
  \bibinfo{person}{Stephen Twigg}, \bibinfo{person}{Huy Vo}, {and}
  \bibinfo{person}{Andrew Waterman}.} \bibinfo{year}{2016}\natexlab{}.
\newblock \showarticletitle{{The Rocket Chip Generator}}.
\newblock \bibinfo{journal}{\emph{EECS Department, University of California,
  Berkeley, Technical Report}} (\bibinfo{year}{2016}).
\newblock


\bibitem[\protect\citeauthoryear{Bahmani, Brasser, Dessouky, Jauernig, Klimmek,
  Sadeghi, and Stapf}{Bahmani et~al\mbox{.}}{2020}]%
        {DBLP:journals/corr/abs-2010-15866}
\bibfield{author}{\bibinfo{person}{Raad Bahmani}, \bibinfo{person}{Ferdinand
  Brasser}, \bibinfo{person}{Ghada Dessouky}, \bibinfo{person}{Patrick
  Jauernig}, \bibinfo{person}{Matthias Klimmek}, \bibinfo{person}{Ahmad{-}Reza
  Sadeghi}, {and} \bibinfo{person}{Emmanuel Stapf}.}
  \bibinfo{year}{2020}\natexlab{}.
\newblock \showarticletitle{\href{https://arxiv.org/abs/2010.15866}{{CURE:} {A}
  Security Architecture with CUstomizable and Resilient Enclaves}}.
\newblock \bibinfo{journal}{\emph{arXiv abs/2010.15866}}
  (\bibinfo{year}{2020}).
\newblock


\bibitem[\protect\citeauthoryear{Beniamini}{Beniamini}{2016a}]%
        {Bits2016}
\bibfield{author}{\bibinfo{person}{Gal Beniamini}.}
  \bibinfo{year}{2016}\natexlab{a}.
\newblock
  \bibinfo{booktitle}{\emph{\href{http://bits-please.blogspot.com/2016/05/qsee-privilege-escalation-vulnerability.html}{QSEE
  privilege escalation vulnerability and exploit (CVE-2015-6639)}}}.
\newblock


\bibitem[\protect\citeauthoryear{Beniamini}{Beniamini}{2016b}]%
        {WoW2016}
\bibfield{author}{\bibinfo{person}{Gal Beniamini}.}
  \bibinfo{year}{2016}\natexlab{b}.
\newblock
  \bibinfo{booktitle}{\emph{\href{https://bits-please.blogspot.com/2016/05/war-of-worlds-hijacking-linux-kernel.html}{War
  of the Worlds - Hijacking the Linux Kernel from QSEE}}}.
\newblock


\bibitem[\protect\citeauthoryear{Bhartiya}{Bhartiya}{2020}]%
        {LinuxLOC}
\bibfield{author}{\bibinfo{person}{Swapnil Bhartiya}.}
  \bibinfo{year}{2020}\natexlab{}.
\newblock
  \bibinfo{booktitle}{\emph{\href{https://www.linux.com/news/linux-in-2020-27-8-million-lines-of-code-in-the-kernel-1-3-million-in-systemd/}{Linux
  in 2020: 27.8 million lines of code in the kernel, 1.3 million in systemd}}}.
\newblock


\bibitem[\protect\citeauthoryear{Brasser, Gens, Jauernig, Sadeghi, and
  Stapf}{Brasser et~al\mbox{.}}{2019}]%
        {DBLP:conf/ndss/BrasserGJSS19}
\bibfield{author}{\bibinfo{person}{Ferdinand Brasser}, \bibinfo{person}{David
  Gens}, \bibinfo{person}{Patrick Jauernig}, \bibinfo{person}{Ahmad{-}Reza
  Sadeghi}, {and} \bibinfo{person}{Emmanuel Stapf}.}
  \bibinfo{year}{2019}\natexlab{}.
\newblock
  \showarticletitle{\href{https://www.ndss-symposium.org/ndss-paper/sanctuary-arming-trustzone-with-user-space-enclaves/}{{SANCTUARY:}
  ARMing TrustZone with User-space Enclaves}}. In
  \bibinfo{booktitle}{\emph{Network and Distributed System Security Symposium
  -- {NDSS}}}.
\newblock


\bibitem[\protect\citeauthoryear{Brasser, M{\"{u}}ller, Dmitrienko, Kostiainen,
  Capkun, and Sadeghi}{Brasser et~al\mbox{.}}{2017}]%
        {DBLP:conf/woot/BrasserMDKCS17}
\bibfield{author}{\bibinfo{person}{Ferdinand Brasser}, \bibinfo{person}{Urs
  M{\"{u}}ller}, \bibinfo{person}{Alexandra Dmitrienko}, \bibinfo{person}{Kari
  Kostiainen}, \bibinfo{person}{Srdjan Capkun}, {and}
  \bibinfo{person}{Ahmad{-}Reza Sadeghi}.} \bibinfo{year}{2017}\natexlab{}.
\newblock
  \showarticletitle{\href{https://www.usenix.org/conference/woot17/workshop-program/presentation/brasser}{Software
  Grand Exposure: {SGX} Cache Attacks Are Practical}}. In
  \bibinfo{booktitle}{\emph{Workshop on Offensive Technologies -- {WOOT}}}.
\newblock


\bibitem[\protect\citeauthoryear{Bulck, Minkin, Weisse, Genkin, Kasikci,
  Piessens, Silberstein, Wenisch, Yarom, and Strackx}{Bulck
  et~al\mbox{.}}{2018}]%
        {DBLP:conf/uss/BulckMWGKPSWYS18}
\bibfield{author}{\bibinfo{person}{Jo~Van Bulck}, \bibinfo{person}{Marina
  Minkin}, \bibinfo{person}{Ofir Weisse}, \bibinfo{person}{Daniel Genkin},
  \bibinfo{person}{Baris Kasikci}, \bibinfo{person}{Frank Piessens},
  \bibinfo{person}{Mark Silberstein}, \bibinfo{person}{Thomas~F. Wenisch},
  \bibinfo{person}{Yuval Yarom}, {and} \bibinfo{person}{Raoul Strackx}.}
  \bibinfo{year}{2018}\natexlab{}.
\newblock
  \showarticletitle{\href{https://www.usenix.org/conference/usenixsecurity18/presentation/bulck}{Foreshadow:
  Extracting the Keys to the Intel {SGX} Kingdom with Transient Out-of-Order
  Execution}}. In \bibinfo{booktitle}{\emph{{USENIX} Security Symposium}}.
\newblock


\bibitem[\protect\citeauthoryear{Cerdeira, Santos, Fonseca, and Pinto}{Cerdeira
  et~al\mbox{.}}{2020}]%
        {cerdeira2020sok}
\bibfield{author}{\bibinfo{person}{David Cerdeira}, \bibinfo{person}{Nuno
  Santos}, \bibinfo{person}{Pedro Fonseca}, {and} \bibinfo{person}{Sandro
  Pinto}.} \bibinfo{year}{2020}\natexlab{}.
\newblock \showarticletitle{SoK: Understanding the Prevailing Security
  Vulnerabilities in TrustZone-assisted TEE Systems}. In
  \bibinfo{booktitle}{\emph{Proceedings of the IEEE Symposium on Security and
  Privacy (S\&P), San Francisco, CA, USA}}.
\newblock


\bibitem[\protect\citeauthoryear{Checkoway, Davi, Dmitrienko, Sadeghi, Shacham,
  and Winandy}{Checkoway et~al\mbox{.}}{2010}]%
        {DBLP:conf/ccs/CheckowayDDSSW10}
\bibfield{author}{\bibinfo{person}{Stephen Checkoway}, \bibinfo{person}{Lucas
  Davi}, \bibinfo{person}{Alexandra Dmitrienko}, \bibinfo{person}{Ahmad{-}Reza
  Sadeghi}, \bibinfo{person}{Hovav Shacham}, {and} \bibinfo{person}{Marcel
  Winandy}.} \bibinfo{year}{2010}\natexlab{}.
\newblock
  \showarticletitle{\href{https://doi.org/10.1145/1866307.1866370}{Return-oriented
  programming without returns}}. In \bibinfo{booktitle}{\emph{Conference on
  Computer and Communications Security -- {CCS}}}.
\newblock


\bibitem[\protect\citeauthoryear{Chen, Chen, Xiao, Zhang, Lin, and Lai}{Chen
  et~al\mbox{.}}{2018}]%
        {DBLP:journals/corr/abs-1802-09085}
\bibfield{author}{\bibinfo{person}{Guoxing Chen}, \bibinfo{person}{Sanchuan
  Chen}, \bibinfo{person}{Yuan Xiao}, \bibinfo{person}{Yinqian Zhang},
  \bibinfo{person}{Zhiqiang Lin}, {and} \bibinfo{person}{Ten{-}Hwang Lai}.}
  \bibinfo{year}{2018}\natexlab{}.
\newblock \showarticletitle{\href{http://arxiv.org/abs/1802.09085}{SgxPectre
  Attacks: Leaking Enclave Secrets via Speculative Execution}}.
\newblock \bibinfo{journal}{\emph{arXiv abs/1802.09085}}
  (\bibinfo{year}{2018}).
\newblock


\bibitem[\protect\citeauthoryear{Chen, Chen, Xiao, Zhang, Lin, and Lai}{Chen
  et~al\mbox{.}}{2020}]%
        {DBLP:journals/ieeesp/ChenCXZLL20}
\bibfield{author}{\bibinfo{person}{Guoxing Chen}, \bibinfo{person}{Sanchuan
  Chen}, \bibinfo{person}{Yuan Xiao}, \bibinfo{person}{Yinqian Zhang},
  \bibinfo{person}{Zhiqiang Lin}, {and} \bibinfo{person}{Ten{-}Hwang Lai}.}
  \bibinfo{year}{2020}\natexlab{}.
\newblock
  \showarticletitle{\href{https://doi.org/10.1109/MSEC.2019.2963021}{SgxPectre:
  Stealing Intel Secrets From {SGX} Enclaves via Speculative Execution}}.
\newblock \bibinfo{journal}{\emph{{IEEE} Secur. Priv.}}  \bibinfo{volume}{18}
  (\bibinfo{year}{2020}).
\newblock


\bibitem[\protect\citeauthoryear{Christoulakis, Christou, Athanasopoulos, and
  Ioannidis}{Christoulakis et~al\mbox{.}}{2016}]%
        {DBLP:conf/codaspy/ChristoulakisCA16}
\bibfield{author}{\bibinfo{person}{Nick Christoulakis}, \bibinfo{person}{George
  Christou}, \bibinfo{person}{Elias Athanasopoulos}, {and}
  \bibinfo{person}{Sotiris Ioannidis}.} \bibinfo{year}{2016}\natexlab{}.
\newblock
  \showarticletitle{\href{https://doi.org/10.1145/2857705.2857722}{{HCFI:}
  Hardware-enforced Control-Flow Integrity}}. In
  \bibinfo{booktitle}{\emph{Conference on Data and Application Security and
  Privacy -- {CODASPY}}}.
\newblock


\bibitem[\protect\citeauthoryear{Corporation}{Corporation}{2019}]%
        {SGX2019}
\bibfield{author}{\bibinfo{person}{Intel Corporation}.}
  \bibinfo{year}{2019}\natexlab{}.
\newblock \showarticletitle{Intel 64 and IA-32 Architectures Software
  Developer’s Manual}.
\newblock  (\bibinfo{year}{2019}).
\newblock


\bibitem[\protect\citeauthoryear{Costan, Lebedev, and Devadas}{Costan
  et~al\mbox{.}}{2016}]%
        {DBLP:conf/uss/CostanLD16}
\bibfield{author}{\bibinfo{person}{Victor Costan}, \bibinfo{person}{Ilia~A.
  Lebedev}, {and} \bibinfo{person}{Srinivas Devadas}.}
  \bibinfo{year}{2016}\natexlab{}.
\newblock
  \showarticletitle{\href{https://www.usenix.org/conference/usenixsecurity16/technical-sessions/presentation/costan}{Sanctum:
  Minimal Hardware Extensions for Strong Software Isolation}}. In
  \bibinfo{booktitle}{\emph{{USENIX} Security Symposium}}.
\newblock


\bibitem[\protect\citeauthoryear{de~Clercq, G{\"{o}}tzfried, {\"{U}}bler,
  Maene, and Verbauwhede}{de~Clercq et~al\mbox{.}}{2017}]%
        {DBLP:journals/compsec/ClercqGUMV17}
\bibfield{author}{\bibinfo{person}{Ruan de Clercq}, \bibinfo{person}{Johannes
  G{\"{o}}tzfried}, \bibinfo{person}{David {\"{U}}bler},
  \bibinfo{person}{Pieter Maene}, {and} \bibinfo{person}{Ingrid Verbauwhede}.}
  \bibinfo{year}{2017}\natexlab{}.
\newblock
  \showarticletitle{\href{https://doi.org/10.1016/j.cose.2017.03.013}{{SOFIA:}
  Software and control flow integrity architecture}}.
\newblock \bibinfo{journal}{\emph{Comput. Secur.}}  \bibinfo{volume}{68}
  (\bibinfo{year}{2017}).
\newblock


\bibitem[\protect\citeauthoryear{Ekberg, Kostiainen, and Asokan}{Ekberg
  et~al\mbox{.}}{2014}]%
        {DBLP:journals/ieeesp/EkbergKA14}
\bibfield{author}{\bibinfo{person}{Jan{-}Erik Ekberg}, \bibinfo{person}{Kari
  Kostiainen}, {and} \bibinfo{person}{N. Asokan}.}
  \bibinfo{year}{2014}\natexlab{}.
\newblock \showarticletitle{\href{https://doi.org/10.1109/MSP.2014.38}{The
  Untapped Potential of Trusted Execution Environments on Mobile Devices}}.
\newblock \bibinfo{journal}{\emph{{IEEE} Secur. Priv.}}  \bibinfo{volume}{12}
  (\bibinfo{year}{2014}).
\newblock


\bibitem[\protect\citeauthoryear{Firmware}{Firmware}{2020}]%
        {OpTee2020}
\bibfield{author}{\bibinfo{person}{Trusted Firmware}.}
  \bibinfo{year}{2020}\natexlab{}.
\newblock \bibinfo{booktitle}{\emph{\href{https://www.op-tee.org/}{OP-TEE}}}.
\newblock


\bibitem[\protect\citeauthoryear{G{\"{o}}tzfried, Eckert, Schinzel, and
  M{\"{u}}ller}{G{\"{o}}tzfried et~al\mbox{.}}{2017}]%
        {DBLP:conf/eurosec/GotzfriedESM17}
\bibfield{author}{\bibinfo{person}{Johannes G{\"{o}}tzfried},
  \bibinfo{person}{Moritz Eckert}, \bibinfo{person}{Sebastian Schinzel}, {and}
  \bibinfo{person}{Tilo M{\"{u}}ller}.} \bibinfo{year}{2017}\natexlab{}.
\newblock
  \showarticletitle{\href{https://doi.org/10.1145/3065913.3065915}{Cache
  Attacks on Intel {SGX}}}. In \bibinfo{booktitle}{\emph{Proceedings of the
  10th European Workshop on Systems Security, {EUROSEC} 2017, Belgrade, Serbia,
  April 23, 2017}}.
\newblock


\bibitem[\protect\citeauthoryear{Guanciale, Nemati, Baumann, and Dam}{Guanciale
  et~al\mbox{.}}{2016}]%
        {DBLP:conf/sp/GuancialeNBD16}
\bibfield{author}{\bibinfo{person}{Roberto Guanciale}, \bibinfo{person}{Hamed
  Nemati}, \bibinfo{person}{Christoph Baumann}, {and} \bibinfo{person}{Mads
  Dam}.} \bibinfo{year}{2016}\natexlab{}.
\newblock \showarticletitle{\href{https://doi.org/10.1109/SP.2016.11}{Cache
  Storage Channels: Alias-Driven Attacks and Verified Countermeasures}}. In
  \bibinfo{booktitle}{\emph{{IEEE} Symposium on Security and Privacy --
  {S{\&}P}}}.
\newblock


\bibitem[\protect\citeauthoryear{Inc}{Inc}{2020}]%
        {seppatent}
\bibfield{author}{\bibinfo{person}{Apple Inc}.}
  \bibinfo{year}{2020}\natexlab{}.
\newblock
  \bibinfo{booktitle}{\emph{\href{http://www.google.com/patents/US8832465}{Security
  enclave processor for a system on a chip}}}.
\newblock
\newblock
\shownote{US8832465B2.}


\bibitem[\protect\citeauthoryear{Intel}{Intel}{2017}]%
        {IntelSidechannel}
\bibfield{author}{\bibinfo{person}{Intel}.} \bibinfo{year}{2017}\natexlab{}.
\newblock
  \bibinfo{booktitle}{\emph{\href{https://software.intel.com/content/www/us/en/develop/articles/intel-sgx-and-side-channels.html}{Intel
  SGX and Side-Channels}}}.
\newblock


\bibitem[\protect\citeauthoryear{Jang, Tang, Kim, Sethumadhavan, and Huh}{Jang
  et~al\mbox{.}}{2019}]%
        {DBLP:conf/asplos/JangTKSH19}
\bibfield{author}{\bibinfo{person}{Insu Jang}, \bibinfo{person}{Adrian Tang},
  \bibinfo{person}{Taehoon Kim}, \bibinfo{person}{Simha Sethumadhavan}, {and}
  \bibinfo{person}{Jaehyuk Huh}.} \bibinfo{year}{2019}\natexlab{}.
\newblock
  \showarticletitle{\href{https://doi.org/10.1145/3297858.3304021}{Heterogeneous
  Isolated Execution for Commodity GPUs}}. In
  \bibinfo{booktitle}{\emph{Architectural Support for Programming Languages and
  Operating Systems -- {ASPLOS}}}.
\newblock


\bibitem[\protect\citeauthoryear{Johnson, Rizzo, Ranganathan, McCune, and
  Ho}{Johnson et~al\mbox{.}}{2018}]%
        {johnson2018titan}
\bibfield{author}{\bibinfo{person}{Scott Johnson}, \bibinfo{person}{Dominic
  Rizzo}, \bibinfo{person}{Parthasarathy Ranganathan}, \bibinfo{person}{Jon
  McCune}, {and} \bibinfo{person}{Richard Ho}.}
  \bibinfo{year}{2018}\natexlab{}.
\newblock \showarticletitle{Titan: enabling a transparent silicon root of trust
  for Cloud}. In \bibinfo{booktitle}{\emph{Hot Chips: A Symposium on High
  Performance Chips}}.
\newblock


\bibitem[\protect\citeauthoryear{Khushu and Gomes}{Khushu and Gomes}{2019}]%
        {DBLP:conf/hotchips/KhushuG19}
\bibfield{author}{\bibinfo{person}{Sanjeev Khushu} {and}
  \bibinfo{person}{Wilfred Gomes}.} \bibinfo{year}{2019}\natexlab{}.
\newblock
  \showarticletitle{\href{https://doi.org/10.1109/HOTCHIPS.2019.8875641}{Lakefield:
  Hybrid cores in 3D Package}}. In \bibinfo{booktitle}{\emph{2019 {IEEE} Hot
  Chips 31 Symposium (HCS), Cupertino, CA, USA, August 18-20, 2019}}.
\newblock


\bibitem[\protect\citeauthoryear{Kostiainen, Dhar, and Capkun}{Kostiainen
  et~al\mbox{.}}{2020}]%
        {DBLP:journals/ieeesp/KostiainenDC20}
\bibfield{author}{\bibinfo{person}{Kari Kostiainen}, \bibinfo{person}{Aritra
  Dhar}, {and} \bibinfo{person}{Srdjan Capkun}.}
  \bibinfo{year}{2020}\natexlab{}.
\newblock
  \showarticletitle{\href{https://doi.org/10.1109/MSEC.2020.2990230}{Dedicated
  Security Chips in the Age of Secure Enclaves}}.
\newblock \bibinfo{journal}{\emph{{IEEE} Secur. Priv.}}  \bibinfo{volume}{18}
  (\bibinfo{year}{2020}).
\newblock


\bibitem[\protect\citeauthoryear{Lee, Kohlbrenner, Shinde, Asanovic, and
  Song}{Lee et~al\mbox{.}}{2020}]%
        {DBLP:conf/eurosys/LeeKSAS20}
\bibfield{author}{\bibinfo{person}{Dayeol Lee}, \bibinfo{person}{David
  Kohlbrenner}, \bibinfo{person}{Shweta Shinde}, \bibinfo{person}{Krste
  Asanovic}, {and} \bibinfo{person}{Dawn Song}.}
  \bibinfo{year}{2020}\natexlab{}.
\newblock
  \showarticletitle{\href{https://doi.org/10.1145/3342195.3387532}{Keystone: an
  open framework for architecting trusted execution environments}}. In
  \bibinfo{booktitle}{\emph{EuroSys '20: Fifteenth EuroSys Conference 2020,
  Heraklion, Greece, April 27-30, 2020}}.
\newblock


\bibitem[\protect\citeauthoryear{Lipp, Gruss, Spreitzer, Maurice, and
  Mangard}{Lipp et~al\mbox{.}}{2016}]%
        {DBLP:conf/uss/LippGSMM16}
\bibfield{author}{\bibinfo{person}{Moritz Lipp}, \bibinfo{person}{Daniel
  Gruss}, \bibinfo{person}{Raphael Spreitzer},
  \bibinfo{person}{Cl{\'{e}}mentine Maurice}, {and} \bibinfo{person}{Stefan
  Mangard}.} \bibinfo{year}{2016}\natexlab{}.
\newblock
  \showarticletitle{\href{https://www.usenix.org/conference/usenixsecurity16/technical-sessions/presentation/lipp}{ARMageddon:
  Cache Attacks on Mobile Devices}}. In \bibinfo{booktitle}{\emph{{USENIX}
  Security Symposium}}.
\newblock


\bibitem[\protect\citeauthoryear{LowRISC}{LowRISC}{2019}]%
        {LowRISC2019}
\bibfield{author}{\bibinfo{person}{LowRISC}.} \bibinfo{year}{2019}\natexlab{}.
\newblock \bibinfo{booktitle}{\emph{\href{https://www.lowrisc.org}{{lowRISC
  Chip}}}}.
\newblock


\bibitem[\protect\citeauthoryear{Mandt, Solnik, and Wang}{Mandt
  et~al\mbox{.}}{2016}]%
        {mandt2016demystifying}
\bibfield{author}{\bibinfo{person}{Tarjei Mandt}, \bibinfo{person}{Mathew
  Solnik}, {and} \bibinfo{person}{David Wang}.}
  \bibinfo{year}{2016}\natexlab{}.
\newblock \showarticletitle{Demystifying the secure enclave processor}.
\newblock \bibinfo{journal}{\emph{Black Hat Las Vegas}} (\bibinfo{year}{2016}).
\newblock


\bibitem[\protect\citeauthoryear{McConnell}{McConnell}{2004}]%
        {mcconnell2004code}
\bibfield{author}{\bibinfo{person}{Steve McConnell}.}
  \bibinfo{year}{2004}\natexlab{}.
\newblock \bibinfo{booktitle}{\emph{Code complete}}.
\newblock \bibinfo{publisher}{Pearson Education}.
\newblock


\bibitem[\protect\citeauthoryear{McLellan}{McLellan}{2020}]%
        {AmdPatent}
\bibfield{author}{\bibinfo{person}{Elliot H.~MednickEdward McLellan}.}
  \bibinfo{year}{2020}\natexlab{}.
\newblock \showarticletitle{Instruction subset implementation for low power
  operation}.
\newblock \bibinfo{journal}{\emph{US10698472B2}} (\bibinfo{year}{2020}).
\newblock


\bibitem[\protect\citeauthoryear{McVoy and Staelin}{McVoy and Staelin}{1996}]%
        {DBLP:conf/usenix/McVoyS96}
\bibfield{author}{\bibinfo{person}{Larry~W. McVoy} {and} \bibinfo{person}{Carl
  Staelin}.} \bibinfo{year}{1996}\natexlab{}.
\newblock \showarticletitle{lmbench: Portable Tools for Performance Analysis}.
  In \bibinfo{booktitle}{\emph{{USENIX} Annual Technical Conference}}.
\newblock


\bibitem[\protect\citeauthoryear{Michael~Schwarz}{Michael~Schwarz}{2019}]%
        {miscflaw}
\bibfield{author}{\bibinfo{person}{Erik~Kraft Michael~Schwarz}.}
  \bibinfo{year}{2019}\natexlab{}.
\newblock \showarticletitle{Are Microarchitectural Attacks still possible on
  Flawless Hardware?}
\newblock \bibinfo{journal}{\emph{RuhrSec}} (\bibinfo{year}{2019}).
\newblock


\bibitem[\protect\citeauthoryear{Murdock, Oswald, Garcia, Van~Bulck, Gruss, and
  Piessens}{Murdock et~al\mbox{.}}{2020}]%
        {murdock2020plundervolt}
\bibfield{author}{\bibinfo{person}{Kit Murdock}, \bibinfo{person}{David
  Oswald}, \bibinfo{person}{Flavio~D Garcia}, \bibinfo{person}{Jo Van~Bulck},
  \bibinfo{person}{Daniel Gruss}, {and} \bibinfo{person}{Frank Piessens}.}
  \bibinfo{year}{2020}\natexlab{}.
\newblock \showarticletitle{Plundervolt: Software-based fault injection attacks
  against Intel SGX}. In \bibinfo{booktitle}{\emph{2020 IEEE Symposium on
  Security and Privacy (SP)}}.
\newblock


\bibitem[\protect\citeauthoryear{Pinto and Santos}{Pinto and Santos}{2019}]%
        {DBLP:journals/csur/PintoS19}
\bibfield{author}{\bibinfo{person}{Sandro Pinto} {and} \bibinfo{person}{Nuno
  Santos}.} \bibinfo{year}{2019}\natexlab{}.
\newblock \showarticletitle{\href{https://doi.org/10.1145/3291047}{Demystifying
  Arm TrustZone: {A} Comprehensive Survey}}.
\newblock \bibinfo{journal}{\emph{{ACM} Comput. Surv.}}  \bibinfo{volume}{51}
  (\bibinfo{year}{2019}).
\newblock


\bibitem[\protect\citeauthoryear{Samsung}{Samsung}{2020}]%
        {ese}
\bibfield{author}{\bibinfo{person}{Samsung}.} \bibinfo{year}{2020}\natexlab{}.
\newblock
  \bibinfo{booktitle}{\emph{\href{https://www.samsung.com/semiconductor/security/ese/}{eSE
  Safeguard against digital attacks}}}.
\newblock


\bibitem[\protect\citeauthoryear{Schaffenrath}{Schaffenrath}{2016}]%
        {schaffenrath}
\bibfield{author}{\bibinfo{person}{David Schaffenrath}.}
  \bibinfo{year}{2016}\natexlab{}.
\newblock \showarticletitle{Fault-Attack Secure Processor Design}.
\newblock \bibinfo{journal}{\emph{Graz University of Technology}}
  (\bibinfo{year}{2016}).
\newblock


\bibitem[\protect\citeauthoryear{Schiavone, Conti, Rossi, Gautschi, Pullini,
  Flamand, and Benini}{Schiavone et~al\mbox{.}}{2017}]%
        {schiavone2017slow}
\bibfield{author}{\bibinfo{person}{Pasquale~Davide Schiavone},
  \bibinfo{person}{Francesco Conti}, \bibinfo{person}{Davide Rossi},
  \bibinfo{person}{Michael Gautschi}, \bibinfo{person}{Antonio Pullini},
  \bibinfo{person}{Eric Flamand}, {and} \bibinfo{person}{Luca Benini}.}
  \bibinfo{year}{2017}\natexlab{}.
\newblock \showarticletitle{Slow and steady wins the race? A comparison of
  ultra-low-power RISC-V cores for Internet-of-Things applications}. In
  \bibinfo{booktitle}{\emph{2017 27th International Symposium on Power and
  Timing Modeling, Optimization and Simulation (PATMOS)}}.
\newblock


\bibitem[\protect\citeauthoryear{Schilling, Werner, Nasahl, and
  Mangard}{Schilling et~al\mbox{.}}{2018}]%
        {DBLP:conf/acsac/SchillingWNM18}
\bibfield{author}{\bibinfo{person}{Robert Schilling}, \bibinfo{person}{Mario
  Werner}, \bibinfo{person}{Pascal Nasahl}, {and} \bibinfo{person}{Stefan
  Mangard}.} \bibinfo{year}{2018}\natexlab{}.
\newblock
  \showarticletitle{\href{https://doi.org/10.1145/3274694.3274728}{Pointing in
  the Right Direction - Securing Memory Accesses in a Faulty World}}. In
  \bibinfo{booktitle}{\emph{Annual Computer Security Applications Conference --
  {ACSAC}}}.
\newblock


\bibitem[\protect\citeauthoryear{Schwarz, Weiser, Gruss, Maurice, and
  Mangard}{Schwarz et~al\mbox{.}}{2017}]%
        {DBLP:conf/dimva/SchwarzWGMM17}
\bibfield{author}{\bibinfo{person}{Michael Schwarz}, \bibinfo{person}{Samuel
  Weiser}, \bibinfo{person}{Daniel Gruss}, \bibinfo{person}{Cl{\'{e}}mentine
  Maurice}, {and} \bibinfo{person}{Stefan Mangard}.}
  \bibinfo{year}{2017}\natexlab{}.
\newblock
  \showarticletitle{\href{https://doi.org/10.1007/978-3-319-60876-1_1}{Malware
  Guard Extension: Using {SGX} to Conceal Cache Attacks}}. In
  \bibinfo{booktitle}{\emph{Detection of Intrusions and Malware {\&}
  Vulnerability Assessment -- {DIMVA}}}.
\newblock


\bibitem[\protect\citeauthoryear{Shacham}{Shacham}{2007}]%
        {DBLP:conf/ccs/Shacham07}
\bibfield{author}{\bibinfo{person}{Hovav Shacham}.}
  \bibinfo{year}{2007}\natexlab{}.
\newblock \showarticletitle{\href{https://doi.org/10.1145/1315245.1315313}{The
  geometry of innocent flesh on the bone: return-into-libc without function
  calls (on the x86)}}. In \bibinfo{booktitle}{\emph{Conference on Computer and
  Communications Security -- {CCS}}}.
\newblock


\bibitem[\protect\citeauthoryear{Silva}{Silva}{2019}]%
        {silva2019arm}
\bibfield{author}{\bibinfo{person}{Jos{\'e} Alberto~Moreira Silva}.}
  \bibinfo{year}{2019}\natexlab{}.
\newblock \showarticletitle{Arm TrustZone: evaluating the diversity of the
  memory subsystem}.
\newblock  (\bibinfo{year}{2019}).
\newblock


\bibitem[\protect\citeauthoryear{Technology}{Technology}{2020}]%
        {GlobalPlatformTEE}
\bibfield{author}{\bibinfo{person}{GlobalPlatform~Device Technology}.}
  \bibinfo{year}{2020}\natexlab{}.
\newblock
  \bibinfo{booktitle}{\emph{\href{https://globalplatform.org/specs-library/?filter-committee=tee}{TEE
  Client API Specification}}}.
\newblock


\bibitem[\protect\citeauthoryear{Tice, Roeder, Collingbourne, Checkoway,
  Erlingsson, Lozano, and Pike}{Tice et~al\mbox{.}}{2014}]%
        {DBLP:conf/uss/TiceRCCELP14}
\bibfield{author}{\bibinfo{person}{Caroline Tice}, \bibinfo{person}{Tom
  Roeder}, \bibinfo{person}{Peter Collingbourne}, \bibinfo{person}{Stephen
  Checkoway}, \bibinfo{person}{{\'{U}}lfar Erlingsson}, \bibinfo{person}{Luis
  Lozano}, {and} \bibinfo{person}{Geoff Pike}.}
  \bibinfo{year}{2014}\natexlab{}.
\newblock
  \showarticletitle{\href{https://www.usenix.org/conference/usenixsecurity14/technical-sessions/presentation/tice}{Enforcing
  Forward-Edge Control-Flow Integrity in {GCC} {\&} {LLVM}}}. In
  \bibinfo{booktitle}{\emph{{USENIX} Security Symposium}}.
\newblock


\bibitem[\protect\citeauthoryear{Traber, Zaruba, Stucki, Pullini, Haugou,
  Flamand, Gurkaynak, and Benini}{Traber et~al\mbox{.}}{2016}]%
        {traber2016pulpino}
\bibfield{author}{\bibinfo{person}{Andreas Traber}, \bibinfo{person}{Florian
  Zaruba}, \bibinfo{person}{Sven Stucki}, \bibinfo{person}{Antonio Pullini},
  \bibinfo{person}{Germain Haugou}, \bibinfo{person}{Eric Flamand},
  \bibinfo{person}{Frank~K Gurkaynak}, {and} \bibinfo{person}{Luca Benini}.}
  \bibinfo{year}{2016}\natexlab{}.
\newblock \showarticletitle{PULPino: A small single-core RISC-V SoC}. In
  \bibinfo{booktitle}{\emph{3rd RISCV Workshop}}.
\newblock


\bibitem[\protect\citeauthoryear{Van~Bulck, Moghimi, Schwarz, Lipp, Minkin,
  Genkin, Yarom, Sunar, Gruss, and Piessens}{Van~Bulck et~al\mbox{.}}{2020}]%
        {van2020lvi}
\bibfield{author}{\bibinfo{person}{Jo Van~Bulck}, \bibinfo{person}{Daniel
  Moghimi}, \bibinfo{person}{Michael Schwarz}, \bibinfo{person}{Moritz Lipp},
  \bibinfo{person}{Marina Minkin}, \bibinfo{person}{Daniel Genkin},
  \bibinfo{person}{Yuval Yarom}, \bibinfo{person}{Berk Sunar},
  \bibinfo{person}{Daniel Gruss}, {and} \bibinfo{person}{Frank Piessens}.}
  \bibinfo{year}{2020}\natexlab{}.
\newblock \showarticletitle{LVI: Hijacking transient execution through
  microarchitectural load value injection}. In \bibinfo{booktitle}{\emph{41th
  IEEE Symposium on Security and Privacy (S\&P’20)}}.
\newblock


\bibitem[\protect\citeauthoryear{Volos, Vaswani, and Bruno}{Volos
  et~al\mbox{.}}{2018}]%
        {DBLP:conf/osdi/VolosVB18}
\bibfield{author}{\bibinfo{person}{Stavros Volos}, \bibinfo{person}{Kapil
  Vaswani}, {and} \bibinfo{person}{Rodrigo Bruno}.}
  \bibinfo{year}{2018}\natexlab{}.
\newblock
  \showarticletitle{\href{https://www.usenix.org/conference/osdi18/presentation/volos}{Graviton:
  Trusted Execution Environments on GPUs}}. In
  \bibinfo{booktitle}{\emph{{USENIX} Symposium on Operating Systems Design and
  Implementation -- {OSDI}}}.
\newblock


\bibitem[\protect\citeauthoryear{Wang and Jiang}{Wang and Jiang}{2010}]%
        {DBLP:conf/sp/WangJ10}
\bibfield{author}{\bibinfo{person}{Zhi Wang} {and} \bibinfo{person}{Xuxian
  Jiang}.} \bibinfo{year}{2010}\natexlab{}.
\newblock
  \showarticletitle{\href{https://doi.org/10.1109/SP.2010.30}{HyperSafe: {A}
  Lightweight Approach to Provide Lifetime Hypervisor Control-Flow Integrity}}.
  In \bibinfo{booktitle}{\emph{{IEEE} Symposium on Security and Privacy --
  {S{\&}P}}}.
\newblock


\bibitem[\protect\citeauthoryear{Waterman, Lee, Patterson, and
  Asanović}{Waterman et~al\mbox{.}}{2011}]%
        {NON:Waterman:EECS-2011-62}
\bibfield{author}{\bibinfo{person}{Andrew Waterman}, \bibinfo{person}{Yunsup
  Lee}, \bibinfo{person}{David~A. Patterson}, {and} \bibinfo{person}{Krste
  Asanović}.} \bibinfo{year}{2011}\natexlab{}.
\newblock
  \bibinfo{booktitle}{\emph{\href{http://www2.eecs.berkeley.edu/Pubs/TechRpts/2011/EECS-2011-62.html}{{The
  RISC-V Instruction Set Manual, Volume I: Base User-Level ISA}}}}.
\newblock \bibinfo{type}{{T}echnical {R}eport}. \bibinfo{institution}{EECS
  Department, University of California, Berkeley}.
\newblock


\bibitem[\protect\citeauthoryear{Werner, Unterluggauer, Schaffenrath, and
  Mangard}{Werner et~al\mbox{.}}{2018}]%
        {DBLP:conf/eurosp/WernerUSM18}
\bibfield{author}{\bibinfo{person}{Mario Werner}, \bibinfo{person}{Thomas
  Unterluggauer}, \bibinfo{person}{David Schaffenrath}, {and}
  \bibinfo{person}{Stefan Mangard}.} \bibinfo{year}{2018}\natexlab{}.
\newblock
  \showarticletitle{\href{https://doi.org/10.1109/EuroSP.2018.00023}{Sponge-Based
  Control-Flow Protection for IoT Devices}}. In
  \bibinfo{booktitle}{\emph{European Symposium on Security and Privacy --
  {EuroS{\&}P}}}.
\newblock


\bibitem[\protect\citeauthoryear{Werner, Wenger, and Mangard}{Werner
  et~al\mbox{.}}{2015}]%
        {DBLP:conf/cardis/WernerWM15}
\bibfield{author}{\bibinfo{person}{Mario Werner}, \bibinfo{person}{Erich
  Wenger}, {and} \bibinfo{person}{Stefan Mangard}.}
  \bibinfo{year}{2015}\natexlab{}.
\newblock
  \showarticletitle{\href{https://doi.org/10.1007/978-3-319-31271-2_10}{Protecting
  the Control Flow of Embedded Processors against Fault Attacks}}. In
  \bibinfo{booktitle}{\emph{Smart Card Research and Advanced Applications --
  {CARDIS}}}.
\newblock


\bibitem[\protect\citeauthoryear{Weston}{Weston}{2020}]%
        {pluton}
\bibfield{author}{\bibinfo{person}{David Weston}.}
  \bibinfo{year}{2020}\natexlab{}.
\newblock
  \bibinfo{booktitle}{\emph{\href{https://www.microsoft.com/security/blog/2020/11/17/meet-the-microsoft-pluton-processor-the-security-chip-designed-for-the-future-of-windows-pcs/}{Meet
  the Microsoft Pluton processor – The security chip designed for the future
  of Windows PCs}}}.
\newblock


\bibitem[\protect\citeauthoryear{Wheeler}{Wheeler}{2019}]%
        {WorldGuard19}
\bibfield{author}{\bibinfo{person}{Bob Wheeler}.}
  \bibinfo{year}{2019}\natexlab{}.
\newblock \showarticletitle{SIFIVE SECURES RISC-V}.
\newblock \bibinfo{journal}{\emph{Microprocessor report}}
  (\bibinfo{year}{2019}).
\newblock


\bibitem[\protect\citeauthoryear{Wistoff, Schneider, G{\"{u}}rkaynak, Benini,
  and Heiser}{Wistoff et~al\mbox{.}}{2020}]%
        {DBLP:journals/corr/abs-2005-02193}
\bibfield{author}{\bibinfo{person}{Nils Wistoff}, \bibinfo{person}{Moritz
  Schneider}, \bibinfo{person}{Frank~K. G{\"{u}}rkaynak}, \bibinfo{person}{Luca
  Benini}, {and} \bibinfo{person}{Gernot Heiser}.}
  \bibinfo{year}{2020}\natexlab{}.
\newblock \showarticletitle{\href{https://arxiv.org/abs/2005.02193}{Prevention
  of Microarchitectural Covert Channels on an Open-Source 64-bit {RISC-V}
  Core}}.
\newblock \bibinfo{journal}{\emph{arXiv abs/2005.02193}}
  (\bibinfo{year}{2020}).
\newblock


\bibitem[\protect\citeauthoryear{Zhang and Sekar}{Zhang and Sekar}{2013}]%
        {DBLP:conf/uss/ZhangS13}
\bibfield{author}{\bibinfo{person}{Mingwei Zhang} {and} \bibinfo{person}{R.
  Sekar}.} \bibinfo{year}{2013}\natexlab{}.
\newblock
  \showarticletitle{\href{https://www.usenix.org/conference/usenixsecurity13/technical-sessions/presentation/Zhang}{Control
  Flow Integrity for {COTS} Binaries}}. In \bibinfo{booktitle}{\emph{{USENIX}
  Security Symposium}}.
\newblock


\bibitem[\protect\citeauthoryear{Zhang, Sun, Sun, Lou, and Hou}{Zhang
  et~al\mbox{.}}{2016}]%
        {DBLP:conf/eurosp/ZhangSSLH16}
\bibfield{author}{\bibinfo{person}{Ning Zhang}, \bibinfo{person}{He Sun},
  \bibinfo{person}{Kun Sun}, \bibinfo{person}{Wenjing Lou}, {and}
  \bibinfo{person}{Yiwei~Thomas Hou}.} \bibinfo{year}{2016}\natexlab{}.
\newblock
  \showarticletitle{\href{https://doi.org/10.1109/EuroSP.2016.34}{CacheKit:
  Evading Memory Introspection Using Cache Incoherence}}. In
  \bibinfo{booktitle}{\emph{European Symposium on Security and Privacy --
  {EuroS{\&}P}}}.
\newblock


\bibitem[\protect\citeauthoryear{Zonenberg and Yener}{Zonenberg and
  Yener}{2016}]%
        {DBLP:conf/ches/ZonenbergY16}
\bibfield{author}{\bibinfo{person}{Andrew~D. Zonenberg} {and}
  \bibinfo{person}{B{\"{u}}lent Yener}.} \bibinfo{year}{2016}\natexlab{}.
\newblock
  \showarticletitle{\href{https://doi.org/10.1007/978-3-662-53140-2_12}{Antikernel:
  {A} Decentralized Secure Hardware-Software Operating System Architecture}}.
  In \bibinfo{booktitle}{\emph{Cryptographic Hardware and Embedded Systems --
  {CHES}}}.
\newblock


\end{thebibliography}

\end{document}